\documentclass[twocolumn,tighten]{aastex61}

\usepackage{textcomp}
\usepackage[flushleft]{threeparttable}
\usepackage{array,multirow}
\usepackage{natbib}
\usepackage{amsmath}
\usepackage{gensymb}
\usepackage{scrextend}
\usepackage{hyperref}

\let\oldFootnote\footnote
\newcommand\nextToken\relax

\renewcommand\footnote[1]{%
    \oldFootnote{#1}\futurelet\nextToken\isFootnote}

\newcommand\isFootnote{%
    \ifx\footnote\nextToken\textsuperscript{,}\fi}

\begin{document}

\title{\textit{Chandra} Observations of NGC 7212: Large-Scale Extended Hard-Xray Emission}

\author{Mackenzie L. Jones}
\affil{Center for Astrophysics $\vert$ Harvard \& Smithsonian, 60 Garden St, Cambridge, MA 02138, USA}

\author{G. Fabbiano}
\affil{Center for Astrophysics $\vert$ Harvard \& Smithsonian, 60 Garden St, Cambridge, MA 02138, USA}

\author{Martin Elvis}
\affil{Center for Astrophysics $\vert$ Harvard \& Smithsonian, 60 Garden St, Cambridge, MA 02138, USA}

\author{A. Paggi}
\affil{INAF-Osservatorio Astrofisico di Torino, Via Osservatorio 20, 10025 Pino Torinese, Italy}

\author{M. Karovska}
\affil{Center for Astrophysics $\vert$ Harvard \& Smithsonian, 60 Garden St, Cambridge, MA 02138, USA}

\author{W. P. Maksym}
\affil{Center for Astrophysics $\vert$ Harvard \& Smithsonian, 60 Garden St, Cambridge, MA 02138, USA}

\author{A. Siemiginowska}
\affil{Center for Astrophysics $\vert$ Harvard \& Smithsonian, 60 Garden St, Cambridge, MA 02138, USA}

\author{J. Raymond}
\affil{Center for Astrophysics $\vert$ Harvard \& Smithsonian, 60 Garden St, Cambridge, MA 02138, USA}

\begin{abstract}

Recent observations of nearby Compton thick (CT) active galactic nuclei (AGNs) with \textit{Chandra} have resolved hard ($>3$ keV) X-ray emission extending out from the central supermassive black hole to kiloparsec scales, challenging the long-held belief that the characteristic hard X-ray continuum and fluorescent Fe K lines originate in the inner $\sim$parsec due to the excitation of obscuring material. 
In this paper we present the results of the most recent \textit{Chandra} ACIS-S observations of NGC 7212, a CT AGN in a compact group of interacting galaxies, with a total effective exposure of $\sim$150 ks.
We find $\sim$20\% of the observed emission is found outside of the central ~kiloparsec, with $\sim$17\% associated with the soft X-rays, and $\sim$3\% with hard X-ray continuum and Fe K line. This emission is extended both along the ionization cone and in the cross-cone direction up to $\sim$3.8 kpc scales. 
The spectrum of NGC 7212 is best represented by a mixture of thermal and photoionization models that indicate the presence of complex gas interactions. These observations are consistent with what is observed in other CT AGN (e.g., ESO 428$-$G014, NGC 1068), providing further evidence that this may be a common phenomenon.
High-resolution observations of extended CT AGN provide an especially valuable environment for understanding how AGN feedback impacts host galaxies on galactic scales.

\end{abstract}


\section{Introduction}\label{sec:intro}

At the center of essentially every massive galaxy is a supermassive black hole (SMBH). 
These SMBHs emit enormous amounts of energy as active galactic nuclei (AGNs) powered by accretion onto the black hole (see \citealt{Kor13,Pad17} for a review). 
In the AGN unified model energy is reflected, transmitted, and absorbed as it propagates out from the central nucleus, leaving traces of the AGN geometry on the observed multiwavelength emission (e.g., \citealt{Law82,Ant93,Urr95,Net15}). 

Until recently, it was believed that the characteristic hard X-ray continuum and fluorescent Fe K lines typical of an AGN could only originate from the excitation of an obscuring material in the inner parsecs. In this classical picture, the central SMBH and accretion disk are closely surrounded by an optically thick, molecular torus-like structure. This torus acts as an efficient screen such that radiation propagates along the opening angle as an ionization cone via direct transmission and reflection off of the obscuring material, while being completely attenuated in the cross-cone plane.
Recent observations, however, have uncovered the presence of hard X-ray emission on $\sim$kiloparsec scales in the direction of the ionization cone (e.g., \textit{Circinus}, \citealt{Are14}; NGC 1068, \citealt{Bau15}; ESO 428$-$G014, \citealt{Fab17}) and cross-cone (e.g., ESO 428$-$G014 \citealt{Fab18II,Fab18III,Fab19}) in nearby Compton thick (CT) AGN. 
The high column densities ($\log\text{N}_{\text{H}} > 23$ cm$^{-2}$) of these CT AGN uniquely allow for these types of investigations as the obscuration depletes the X-ray emission of the central point-like source, revealing the extended material.

In this paper, we present the results of an investigation into the presence of extended hard X-ray emission in NGC 7212,\footnote{NGC 7212 is also called IRAS 22045+0959, CGCG 428-032, MCG +02-56-011, and UGC 11910.} a nearby ($z=0.0266$; $D_{\text{lum}}\sim$$115$ Mpc) Seyfert 2 galaxy with a heavily obscured AGN ($\log M_{\text{bh}}=7.54$; $\log L/L_{Edd}=-1.55$; \citealt{Her15}) located in a compact group of three interacting galaxies (e.g., \citealt{Mun07}). 
NGC 7212 is part of a sample of nearby CT AGN in normal Seyfert 2 galaxies (as classified by their optical emission line ratios) with bright [\ion{O}{3}] $\lambda$5007 cores, and no history of nuclear starbursts (\citealt{Lev06}). 
Other CT AGN in this sample have already been mentioned as exhibiting extended X-ray emission on $\sim$kiloparsec scales (e.g., NGC 1068, \citealt{Bau15}; ESO 428$-$G014, \citealt{Fab17,Fab18II,Fab18III,Fab19}), but NGC 7212 is the most distant of all of these nearby sources and thus is a valuable addition to this recent work in establishing the ubiquity of extended hard X-ray emission.

Previous observations of NGC 7212 have found kiloparsec-scale, diffuse, extended optical narrow line emission (ENLR; e.g., \citealt{Was81,Fal98,Sch03,Cra11,Con17}), and polarized optical broad line emission (e.g., \citealt{Tra95I,Tra95II,Vei97}). It has a compact double radio source (extent $0.7\arcsec$) with moderate radio power aligned with the elongated narrow line emission (e.g., P.A. $\sim -7\degree$, \citealt{Fal98,Dra03}). In the X-rays, NGC 7212 has previously been established as nonvariable with a complex X-ray spectrum exhibiting the characteristic features of a CT AGN (e.g., \citealt{Ris00,Gua05,Bia06,Lev06,Sin11,Sev12,Her15,Mar18}). 

To this extensive multiwavelength coverage, we have added deep X-ray observations of NGC 7212 for a cumulative \textit{Chandra} exposure of 149.87 ks, in order to piece together a detailed picture of the morphological and spectral properties of this CT AGN. 
We describe these observations and the data reduction in Section \ref{sec:obs}, and report on the spatial and spectral properties of the nuclear and extended emission in Sections \ref{sec:rp} and \ref{sec:spec}, respectively. In Section \ref{sec:dis} we discuss the results of the spectral analysis and the implications of an extended hard X-ray component. Our findings and conclusions are summarized in Section \ref{sec:con}.

\section{Observations and Analysis}\label{sec:obs}

\begin{figure*}
\begin{center}
\resizebox{150mm}{!}{\includegraphics{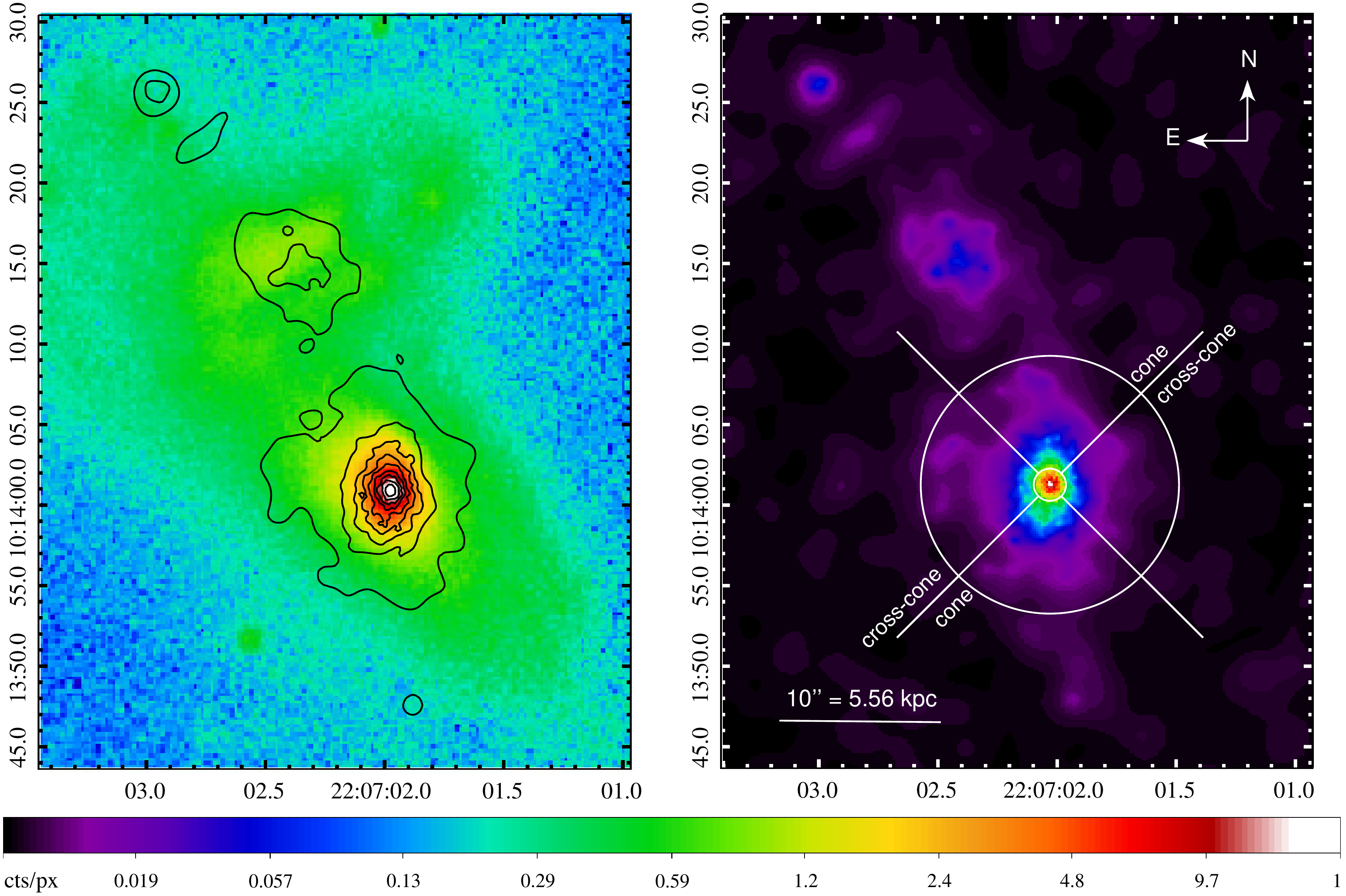}} \\
\caption{Left: X-ray contours (black) corresponding to the merged $0.3-7.0$ keV \textit{Chandra} ACIS image of the three-system interacting galaxy group, including the spiral galaxy NGC 7212 (southwest source), overlaid on a \textit{g}-band Pan-STARRs deep-stack. 
Right: merged $0.3-7.0$ keV \textit{Chandra} ACIS image of NGC 7212 with applied adaptive gaussian smoothing (\textit{dmimgadapt}; $0.5-15$ pixel scales, 5 counts under kernel, 30 iterations) on image pixel $= 1/8$ ACIS pixel. The image contours are logarithmic with colors corresponding to number of counts per image pixel. The nuclear 1\arcsec (0.556 kpc) circular region, and annular 8\arcsec (4.448 kpc) region split into four quadrants, used in the spatial analysis of NGC 7212 are shown in white. \label{fig:img:full}}
\end{center}
\end{figure*}

\begin{table*}
\centering\footnotesize
\caption{Observation Log}\label{tbl:obs}
\scriptsize
\begin{tabular*}{\linewidth}{@{\extracolsep{\fill}}lllllcc}
\hline\hline
ObsID & Instrument & $T_{\text{exp}}$ (ks) & PI & Date & $\delta$x (px) & $\delta$y (px) \\ \hline
4078 & ACIS-S & 19.90 & Kraemer & 2012 Nov 15 & -1.196 & 0.466 \\
20372 & ACIS-S & 49.42 & Fabbiano & 2018 Aug 09 & $-$ & $-$ \\
21668 & ACIS-S & 1.38 & Fabbiano & 2018 Aug 13 & 0.318 & 0.026 \\
21672 & ACIS-S & 27.21 & Fabbiano & 2018 Sep 12 & -1.709 & -0.386 \\ \hline
\end{tabular*}
\end{table*}

We obtained three \textit{Chandra} ACIS-S observations of NGC 7212 (ObsIDs: 20372, 21668, 21672; P.I. Fabbiano) and combined these observations with an additional archival \textit{Chandra} ACIS-S observation of this galaxy (ObsID: 4078; P.I. Kraemer) to generate a dataset with a cumulative effective exposure time of 147.82 ks (Table \ref{tbl:obs}).
These observations were then reprocessed and analyzed using CIAO 4.11 and CALDB 4.8.2 to enable subpixel analysis (\citealt{Tsu01,Wan11}). Each individual observation was inspected for high background flares ($\ge$$3\sigma$) and all were deemed acceptable.

All four observations were exposure corrected and merged, following the CIAO merge threads,\footnote{http://cxc.harvard.edu/ciao/threads/combine/}\footnote{http://cxc.harvard.edu/ciao/threads/merge\_all/} using ObsID 20372 as the reference frame centered at (J2000) RA = 22:07:02.03 (331$\degree$45$\arcmin$30.44$\arcsec$), decl. = +10:14:01.27 (10$\degree$14$\arcmin$1.27$\arcsec$). 
We first visually inspected each observation and determined that manually shifting to the reference image would allow for the best alignment. The final shifts (in native ACIS pixels) are listed in Table \ref{tbl:obs}. The full-band (0.3-7.0 keV) merged image of NGC 7212 and its companion interacting galaxies are shown in Figure \ref{fig:img:full} (right). Contours corresponding to this merged 0.3-7.0 keV image are also shown overlaid on a \textit{g}-band Pan-STARRs deep-stack image in Figure \ref{fig:img:full} (left). We limited our analysis to the 0.3-7.0 keV band, despite typically reliable \textit{Chandra} coverage up to 8.0 keV, due to significant noise.

\section{Spatial Analysis}\label{sec:rp}

\begin{figure*}
\begin{center}
\begin{tabular}{cc}
\resizebox{70mm}{!}{\includegraphics{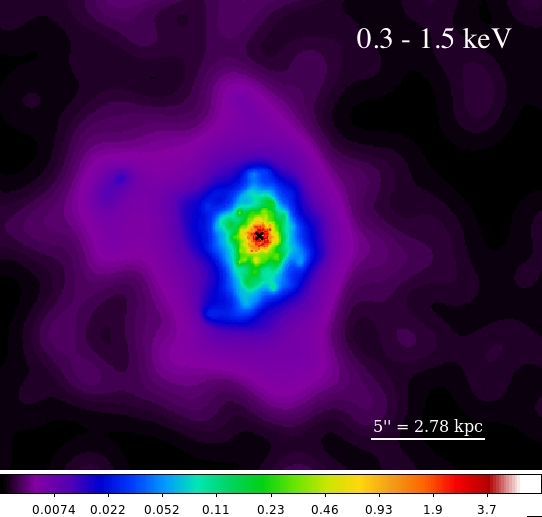}} &  \resizebox{70mm}{!}{\includegraphics{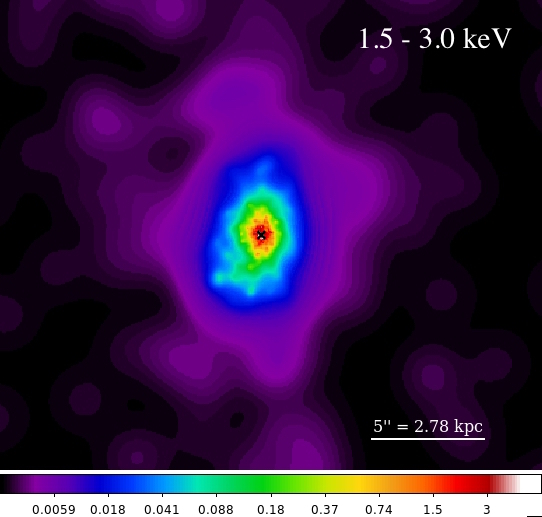}} \\
\resizebox{70mm}{!}{\includegraphics{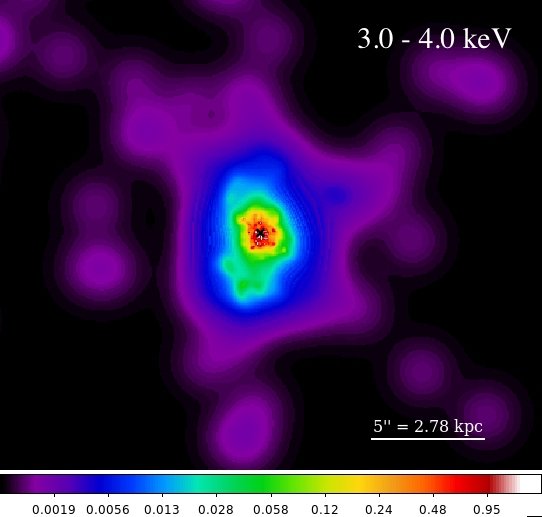}} &  \resizebox{70mm}{!}{\includegraphics{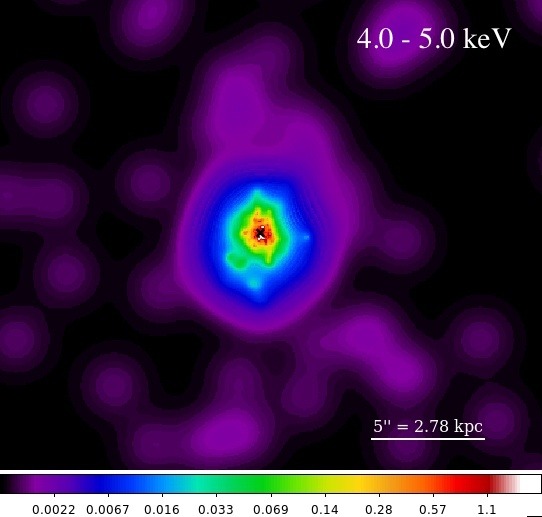}} \\
\resizebox{70mm}{!}{\includegraphics{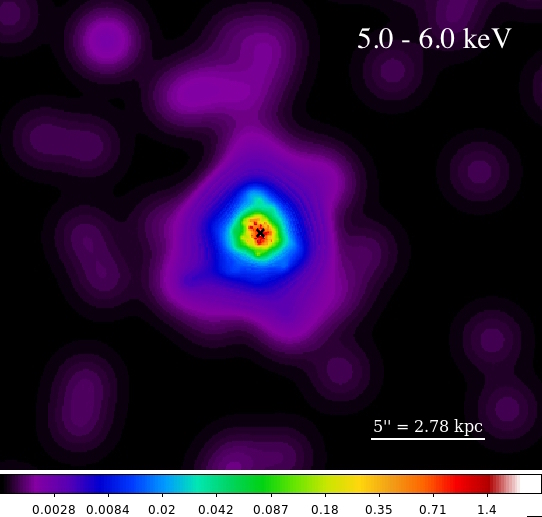}} &  \resizebox{70mm}{!}{\includegraphics{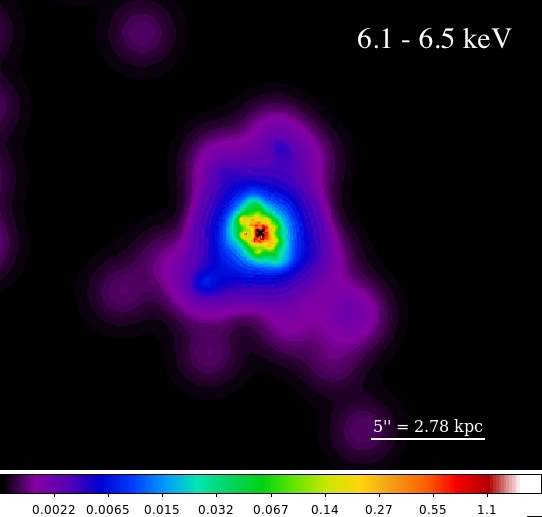}} \\
\end{tabular}
\caption{Adaptively smoothed images of NGC 7212 in the indicated energy bands on image pixel $= 1/8$ ACIS pixel (CIAO adaptive smoothing; $0.5-15$ pixel scales, 5 counts under kernel, 30 iterations). The image contours are logarithmic with colors corresponding to the number of counts per image pixel. The black cross marks the center of the nucleus. \label{fig:img:E}}
\end{center}
\end{figure*}

\begin{figure}
\begin{center}\footnotesize
\begin{tabular}{c}
\resizebox{75mm}{!}{\includegraphics{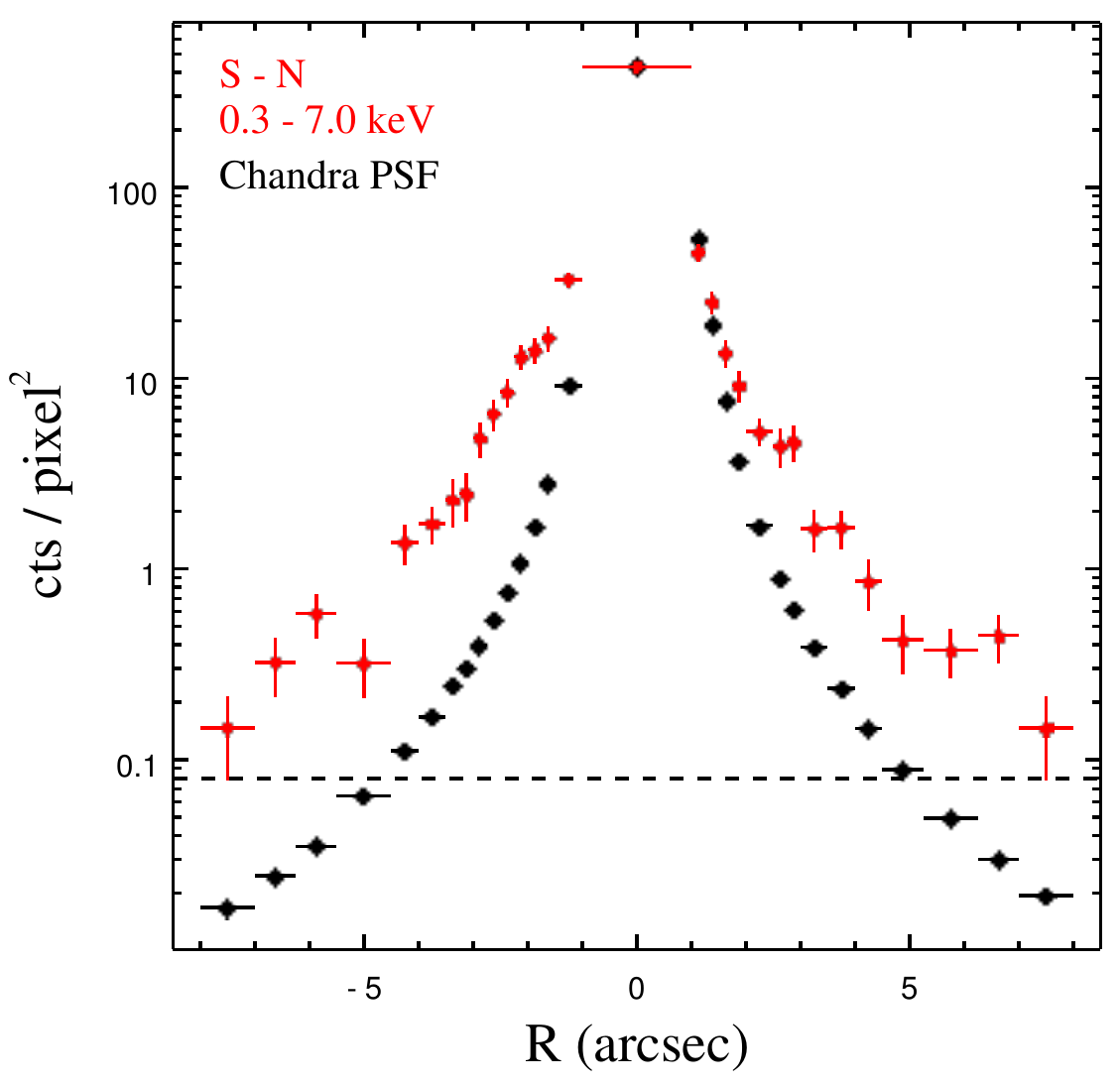}}\\
 \resizebox{75mm}{!}{\includegraphics{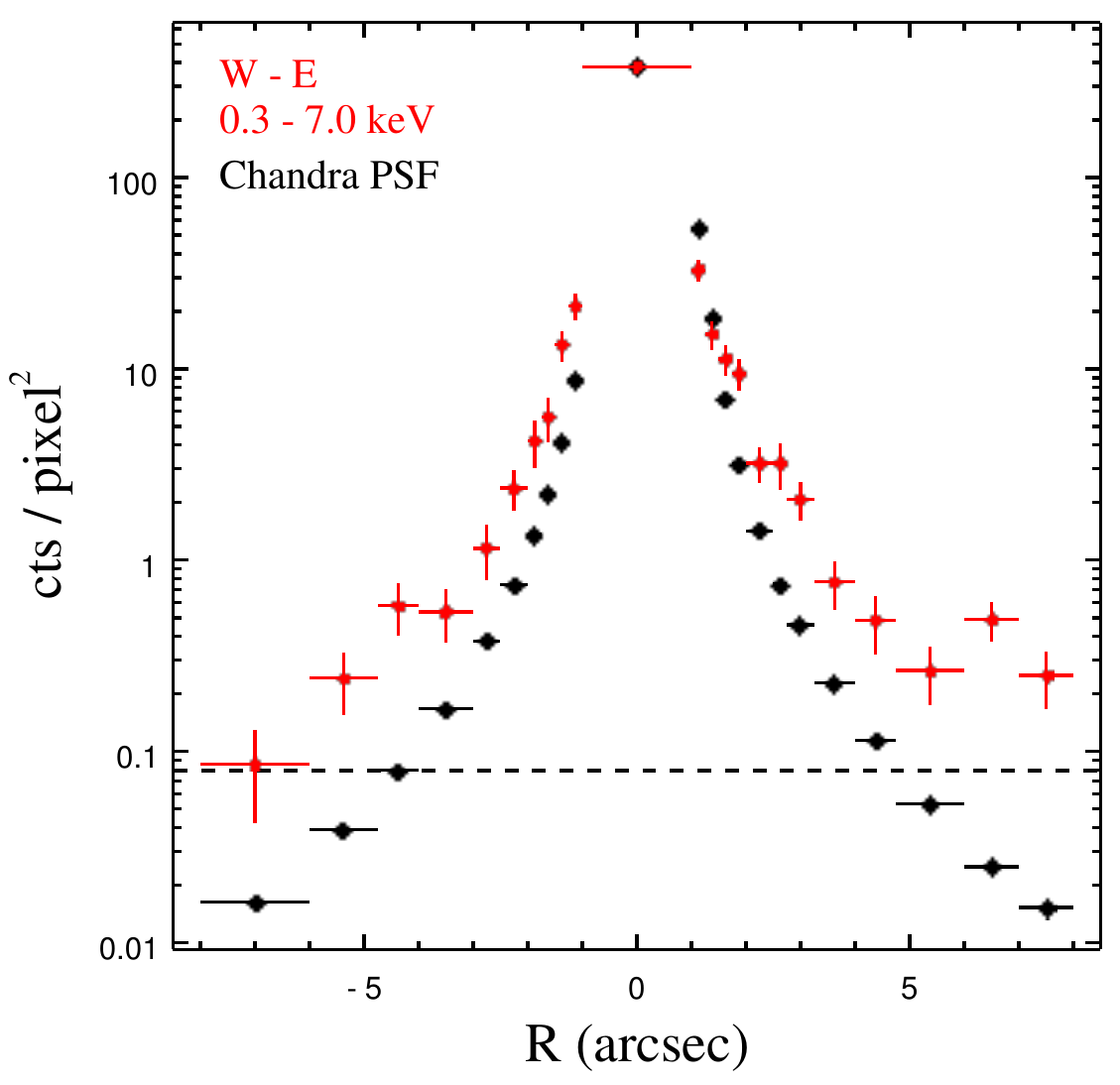}} \\
\end{tabular}
\caption{Background subtracted radial profiles of NGC 7212 for the full energy band ($0.3-7.0$ keV) compared to the \textit{Chandra} PSF which has been renormalized to the 1.0\arcsec~ (0.556 kpc) nuclear region for the (left) south to north and (right) west to east regions. Each bin contains a minimum of 10 counts and is shown with $1\sigma$ errors. We include a dashed horizontal line to indicate the level of background emission and note that points below this line are valid data since the background has already been subtracted from these profiles. We find excess emission observed outside of the central nucleus ($\sim$1.0\arcsec, 0.556 kpc) in each of the given regions. \label{fig:rp:full}}
\end{center}
\end{figure}

\begin{figure*}
\begin{center}\footnotesize
\begin{tabular}{cc}
\resizebox{70mm}{!}{\includegraphics{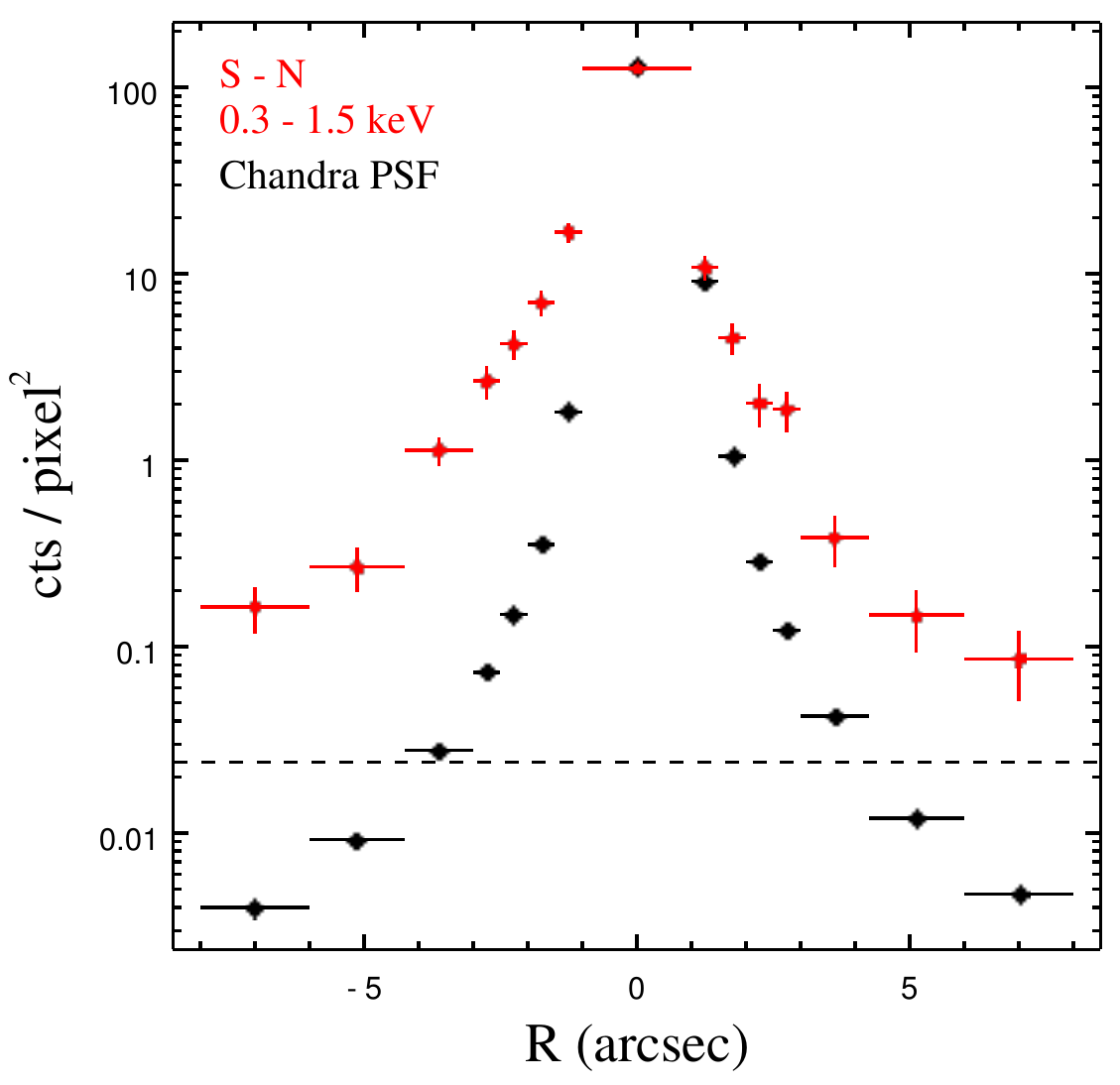}} &  \resizebox{70mm}{!}{\includegraphics{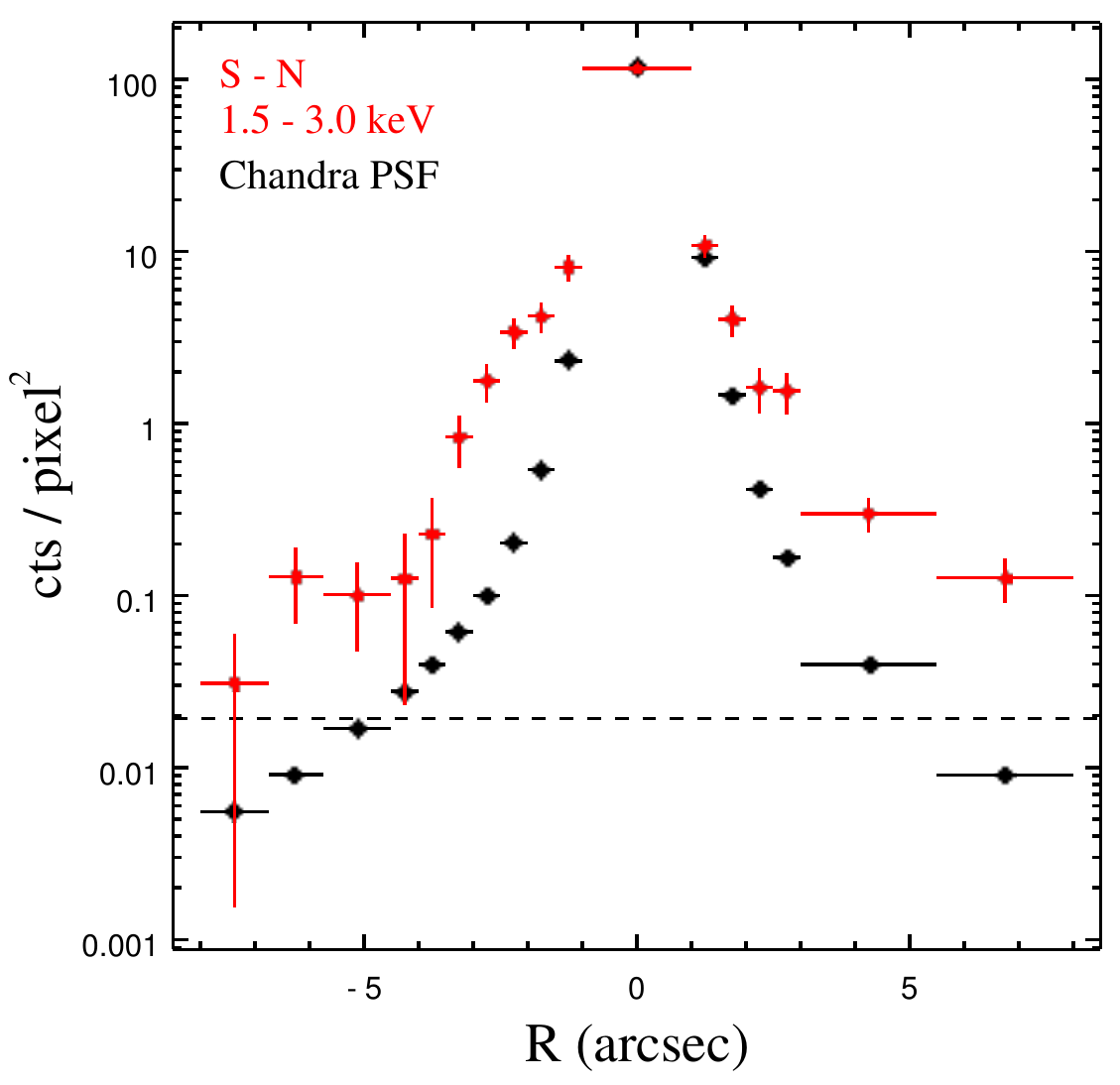}} \\
\resizebox{70mm}{!}{\includegraphics{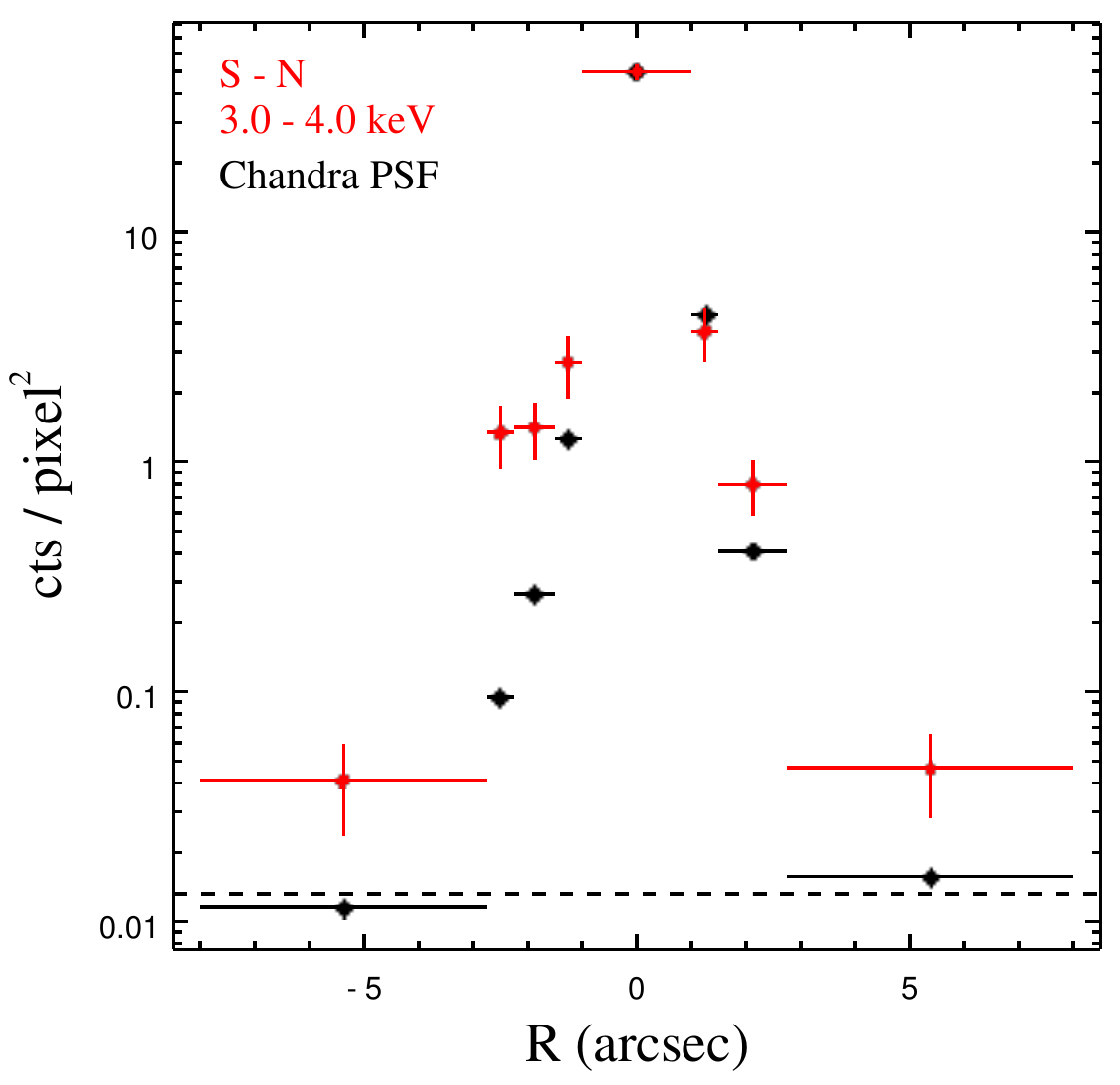}} &  \resizebox{70mm}{!}{\includegraphics{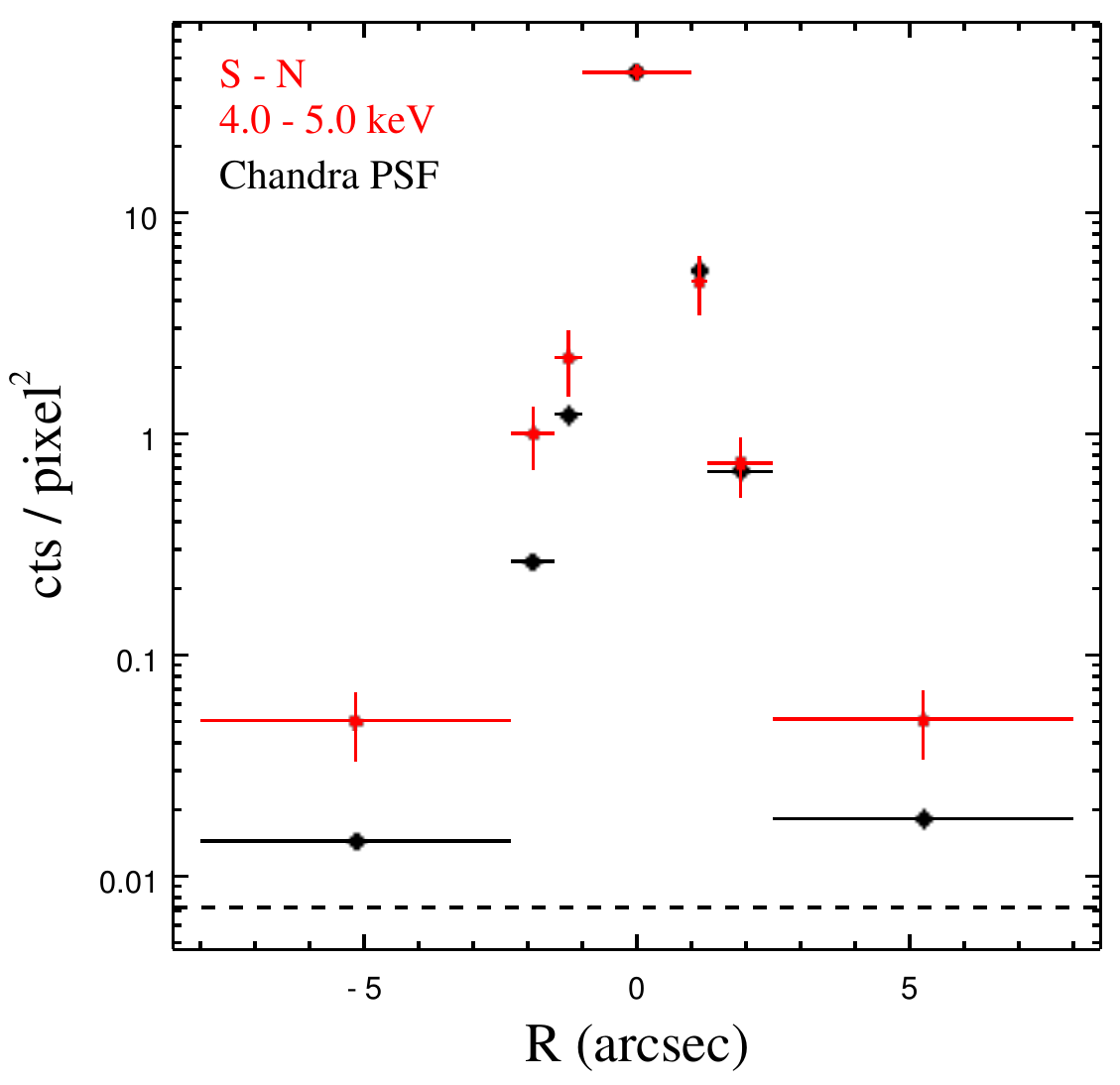}} \\
\resizebox{70mm}{!}{\includegraphics{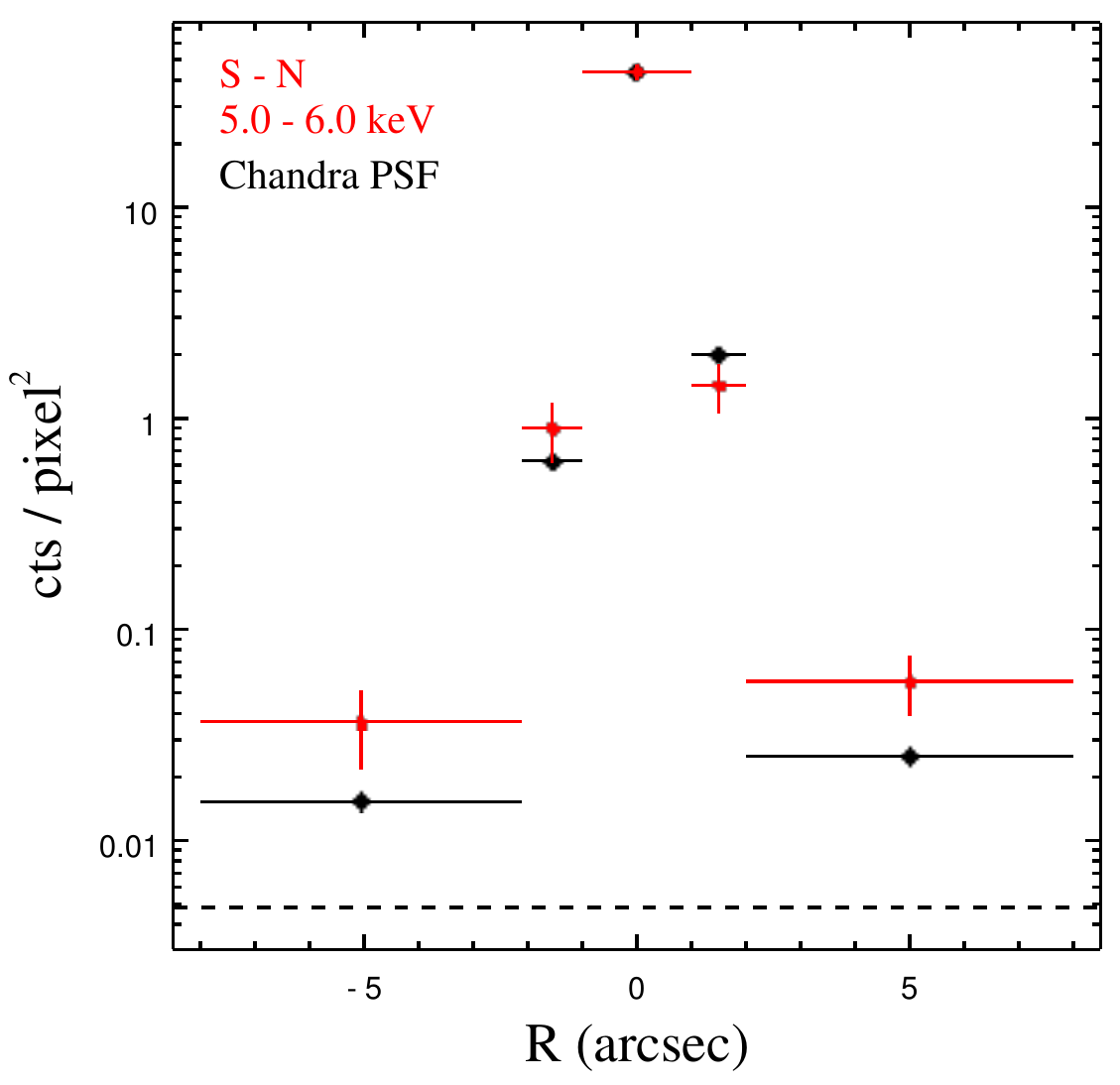}} &  \resizebox{70mm}{!}{\includegraphics{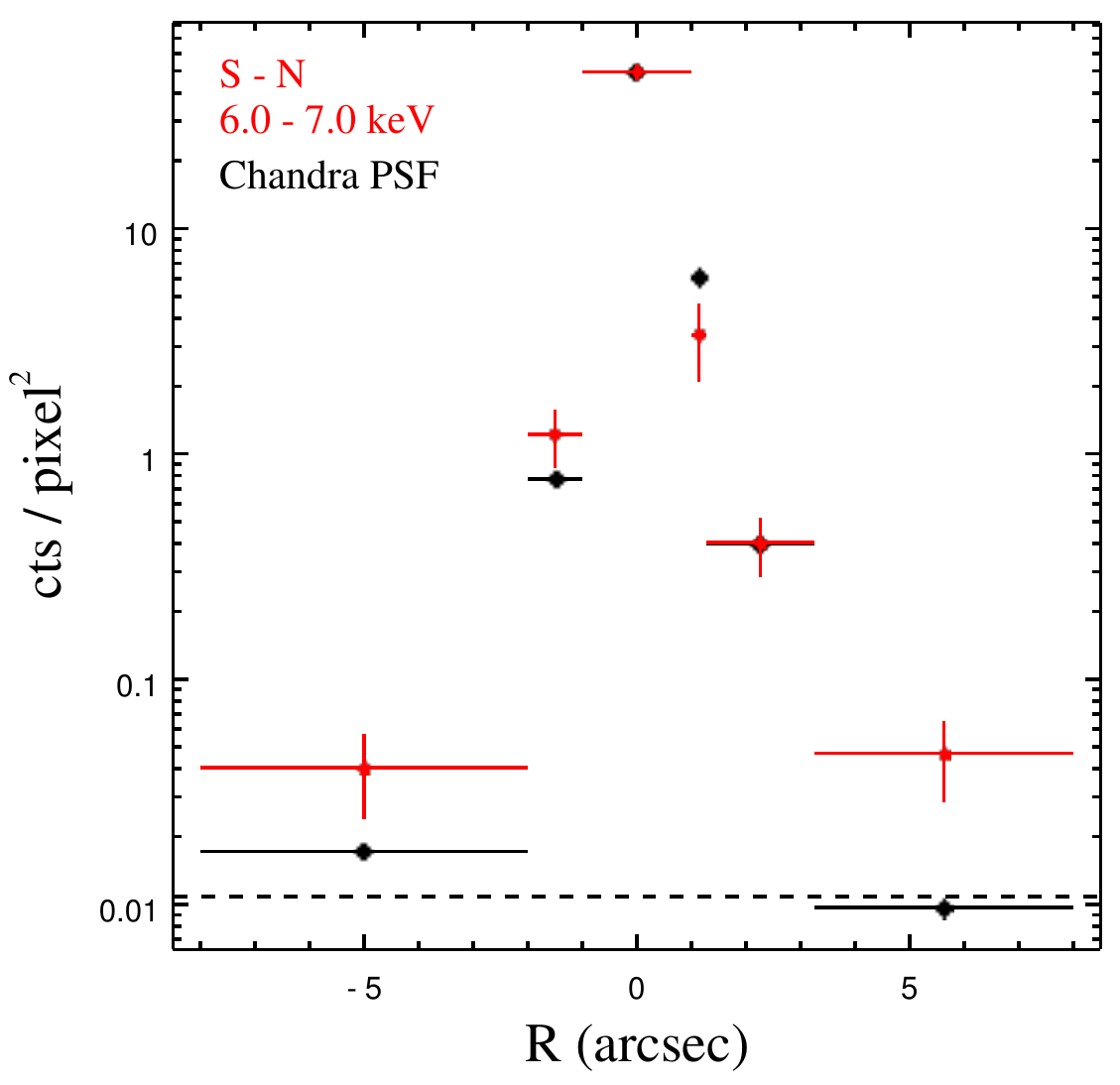}} \\
\end{tabular}
\caption{Similar to Figure \ref{fig:rp:full}. Background subtracted radial profiles of NGC 7212 for the indicated energy bands compared to the \textit{Chandra} PSF which has been renormalized to the 1.0\arcsec~ (0.556 kpc) nuclear region for the south$-$north region. Each bin contains a minimum of 10 counts and is shown with $1\sigma$ errors. We include a dashed horizontal line to indicate the level of background emission and note that points below this line are valid data since the background has already been subtracted from these profiles. \label{fig:rp:sn}}
\end{center}
\end{figure*}

\begin{figure*}
\begin{center}\footnotesize
\begin{tabular}{cc}
\resizebox{70mm}{!}{\includegraphics{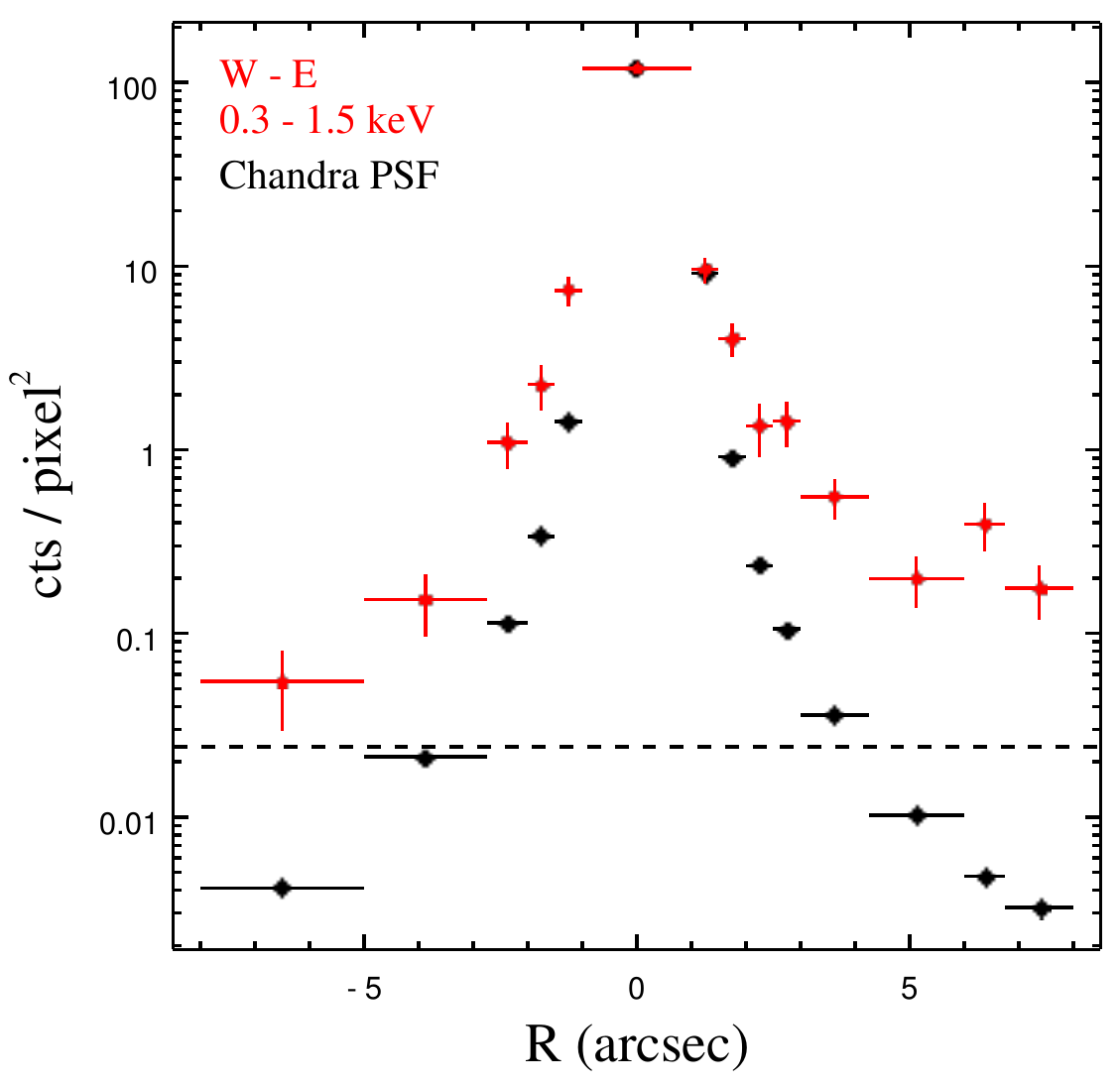}} &  \resizebox{70mm}{!}{\includegraphics{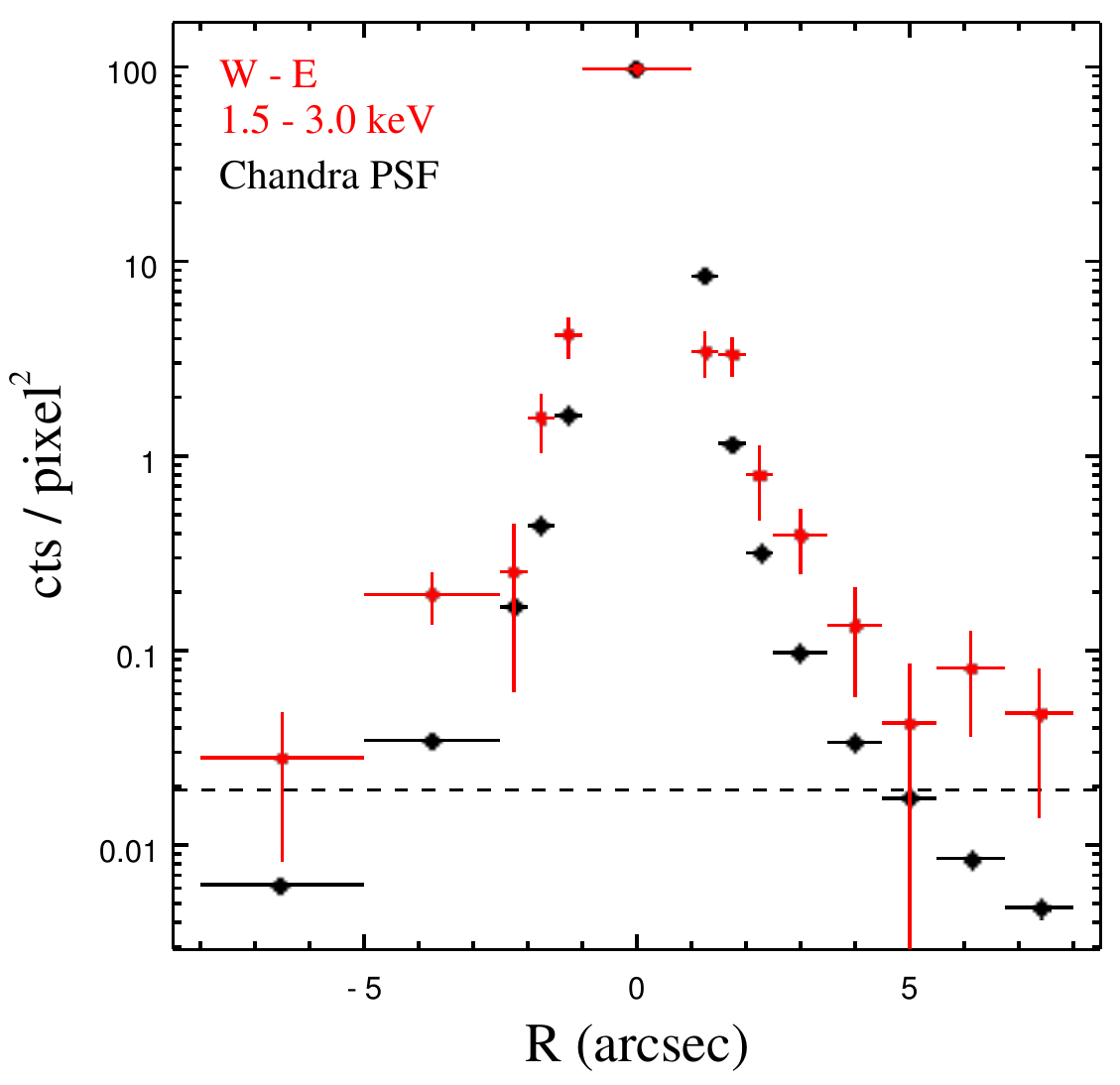}} \\
\resizebox{70mm}{!}{\includegraphics{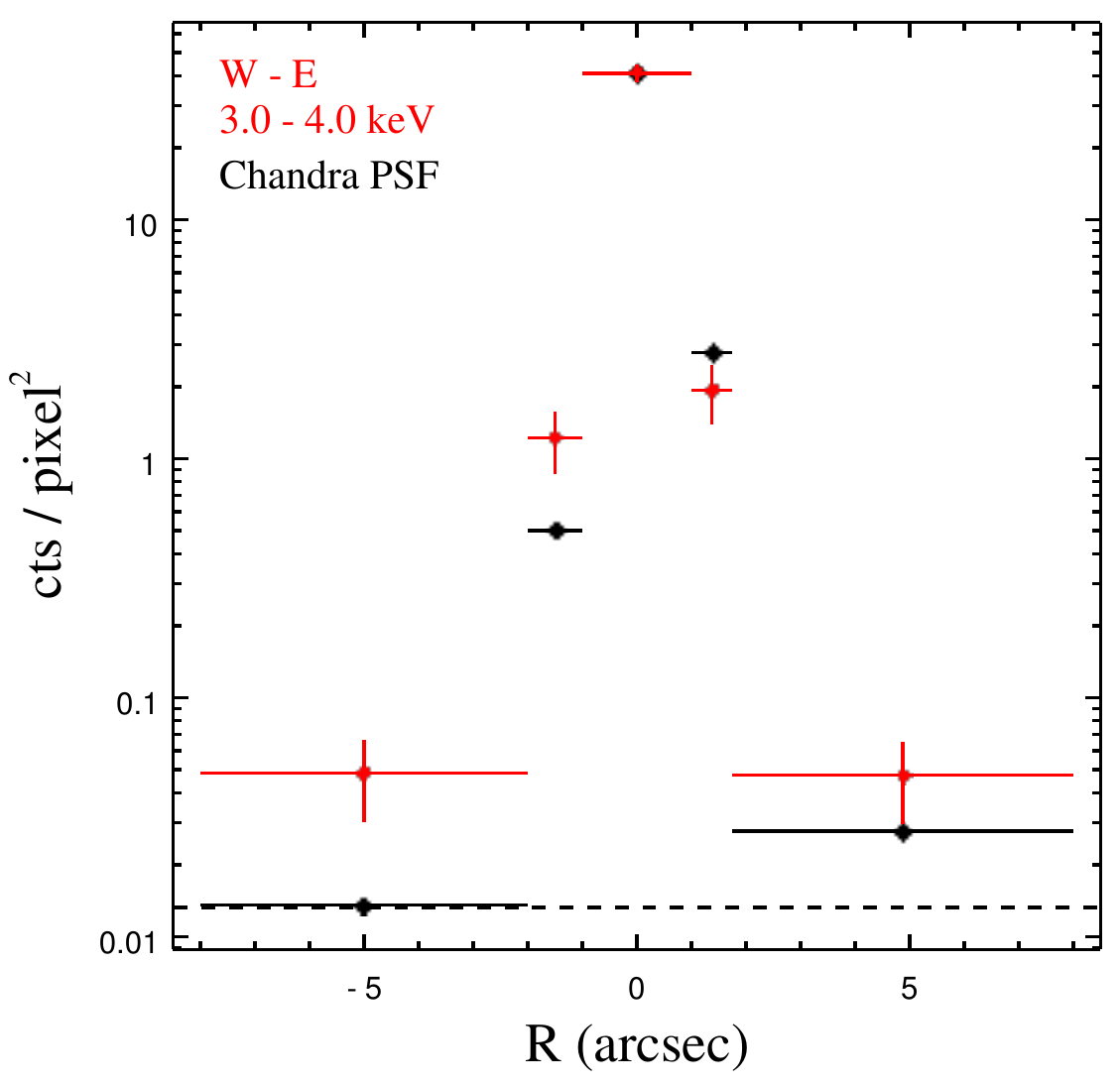}} &  \resizebox{70mm}{!}{\includegraphics{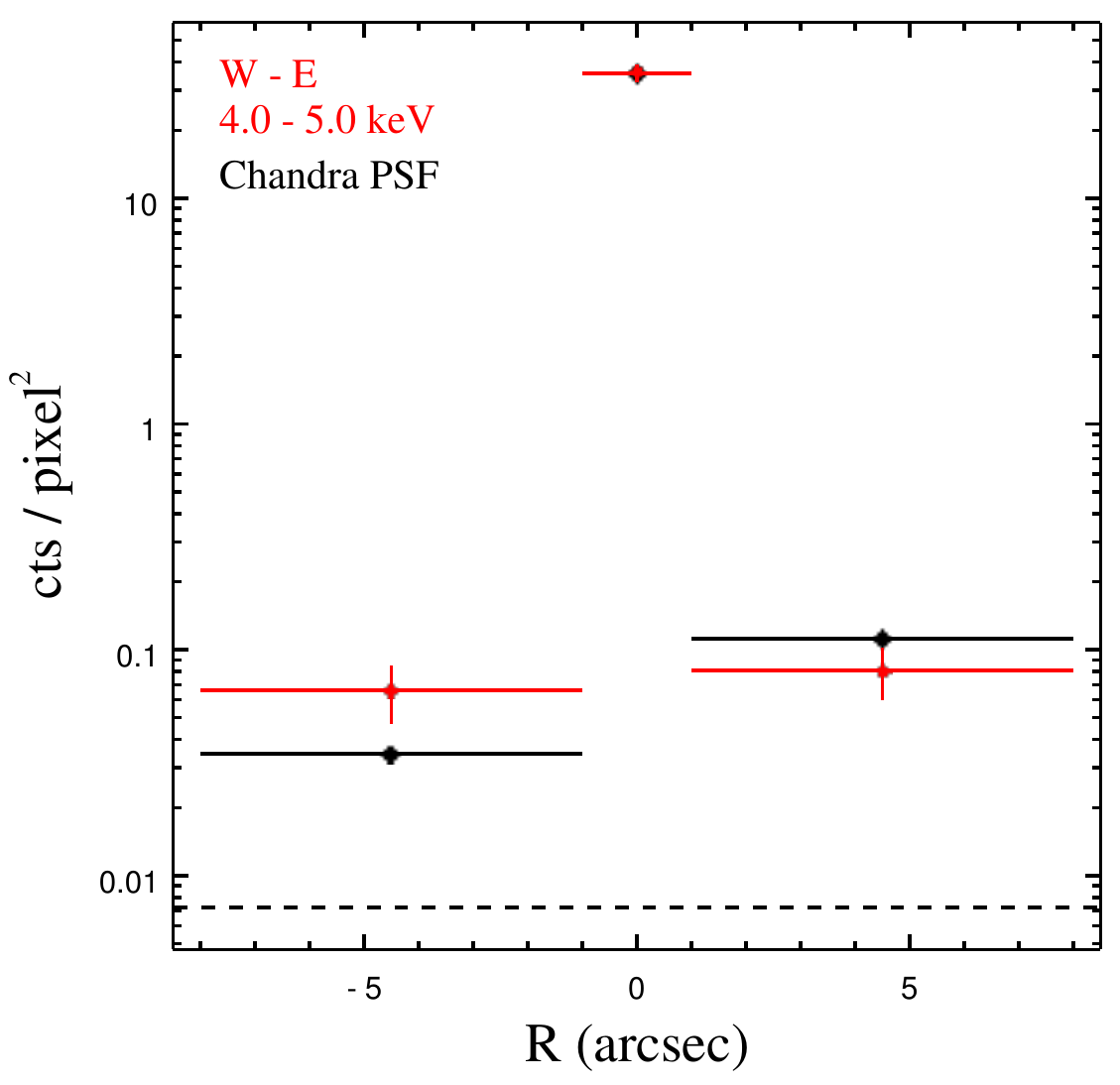}} \\
\resizebox{70mm}{!}{\includegraphics{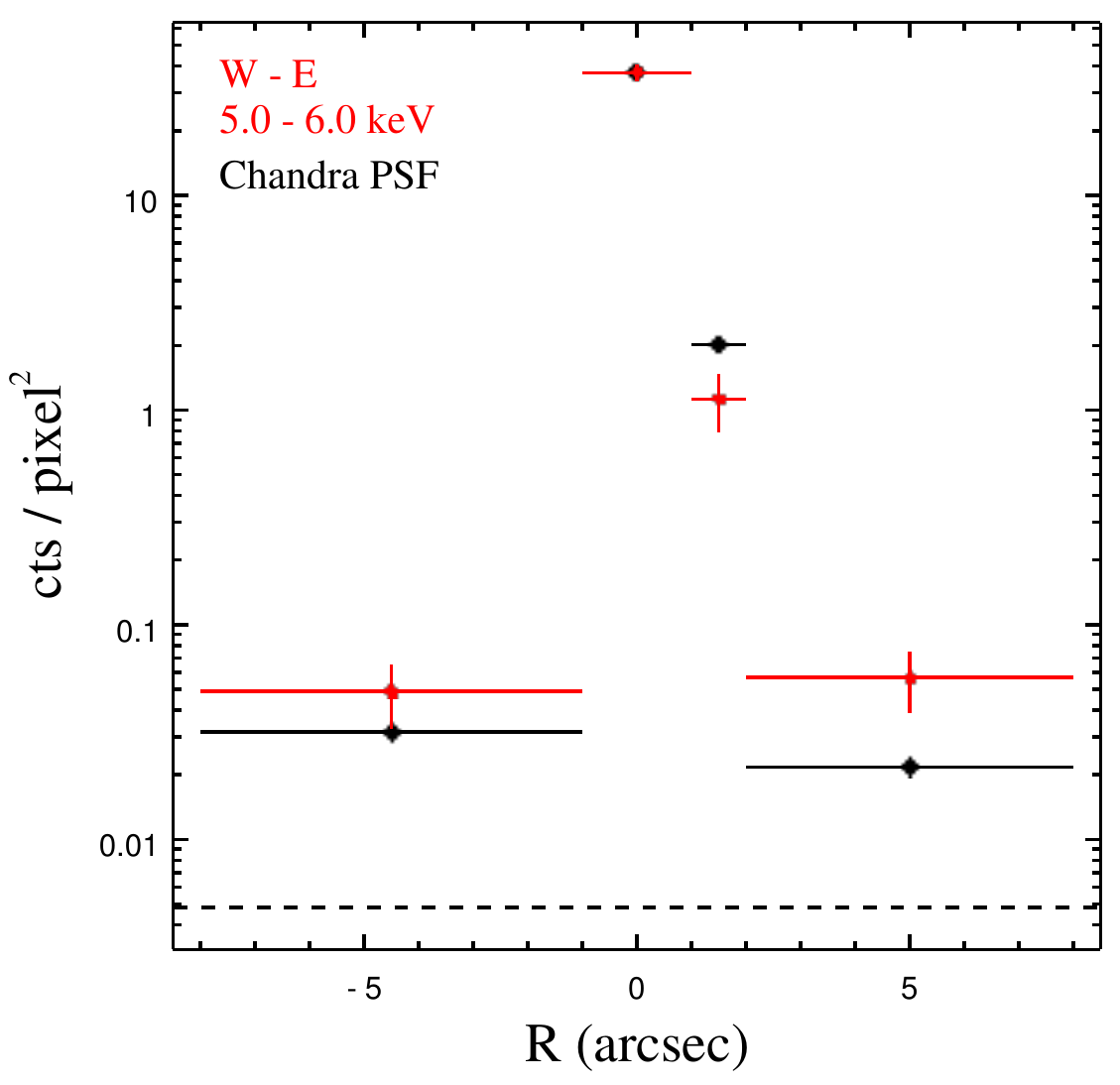}} &  \resizebox{70mm}{!}{\includegraphics{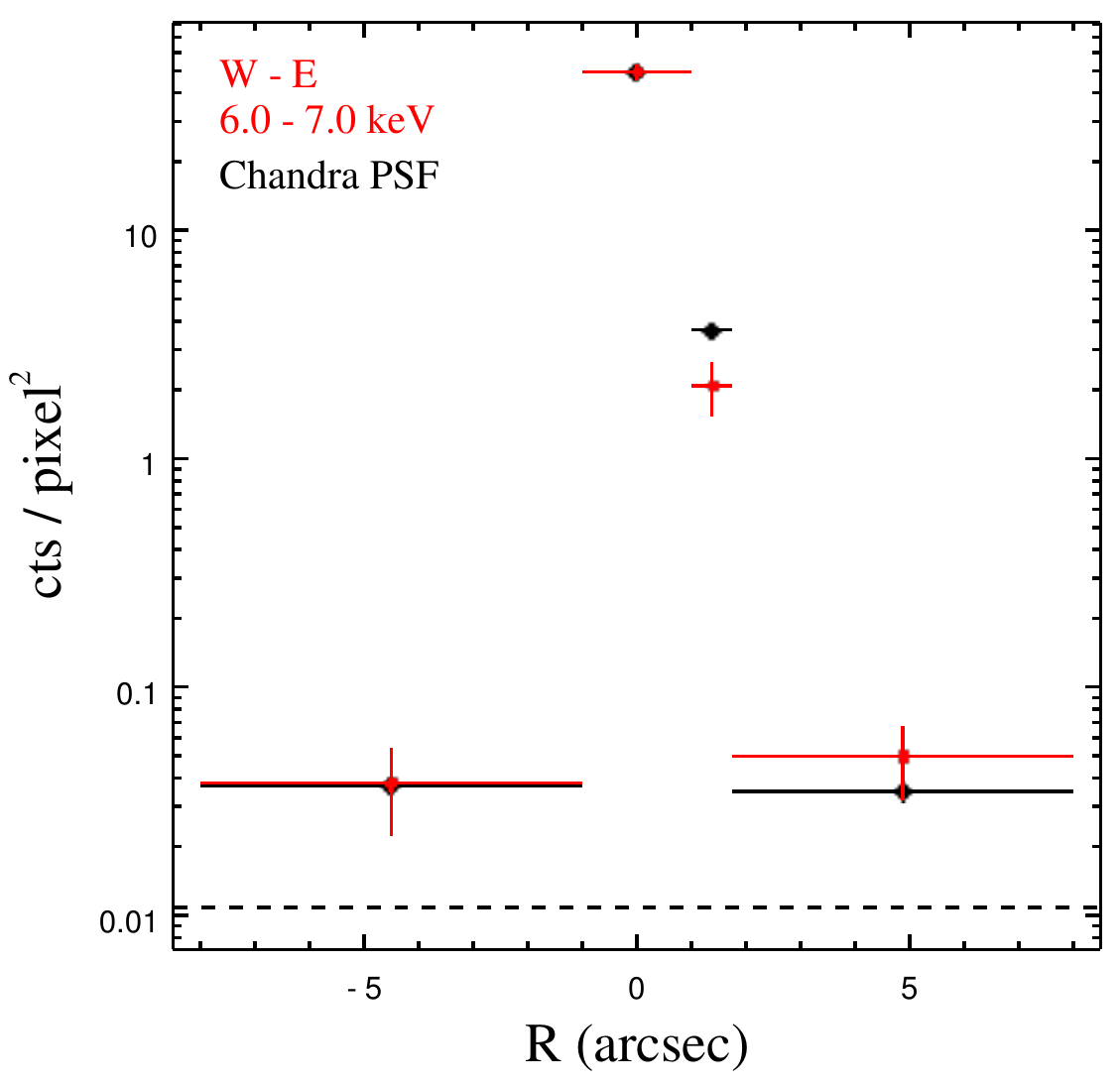}} \\
\end{tabular}
\caption{Same as in Figure \ref{fig:rp:sn}. Background subtracted radial profiles of NGC 7212 for the indicated energy bands compared to the \textit{Chandra} PSF for the west$-$east region. \label{fig:rp:we}}
\end{center}
\end{figure*}

\begin{figure}
\begin{center}
\resizebox{85mm}{!}{\includegraphics{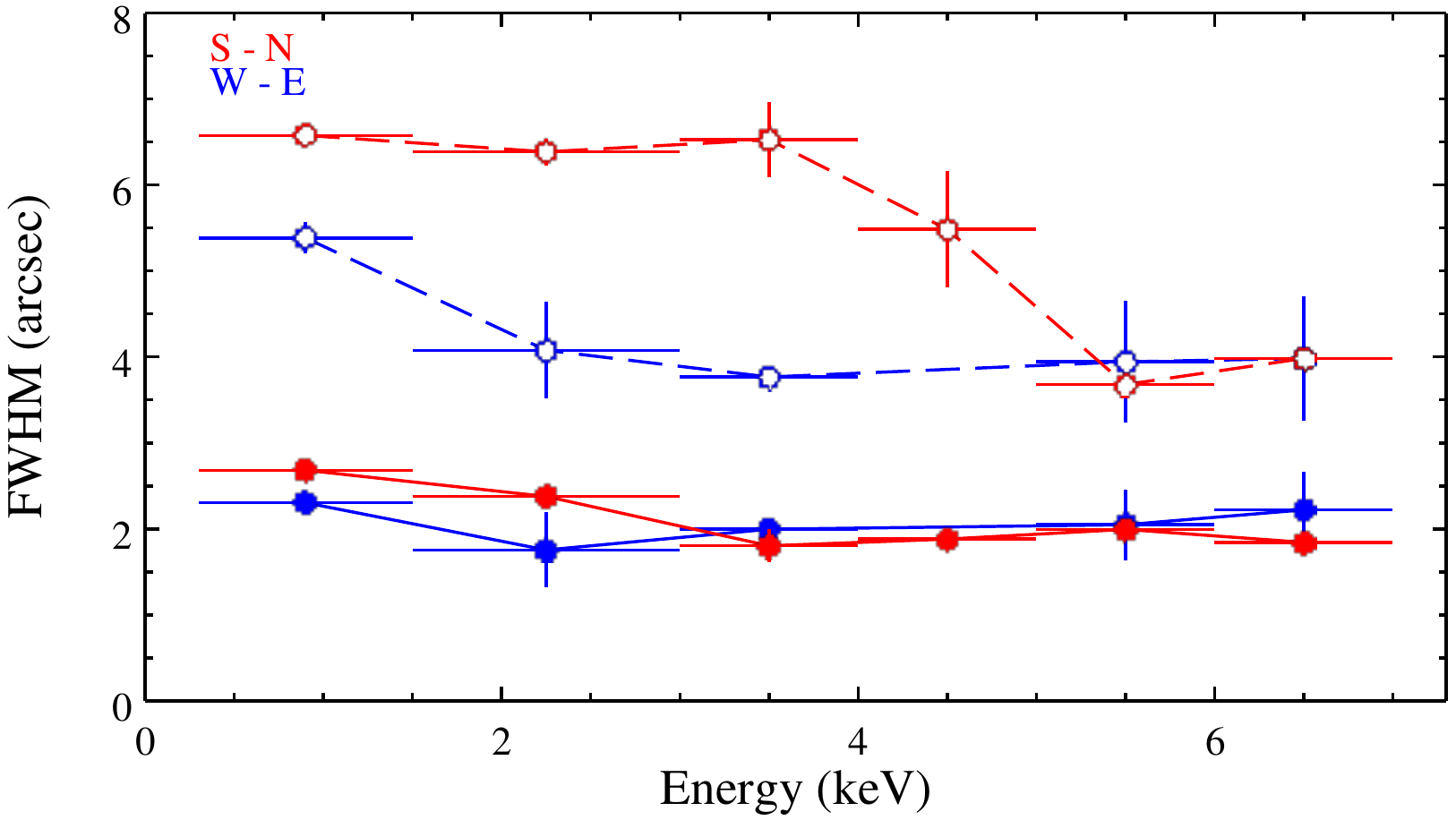}}
\caption{Emission extent as a function of energy calculated from the radial profiles in Figure \ref{fig:rp:sn} and \ref{fig:rp:we} for both the south$-$north and west$-$east regions. Radial profiles with less than four points were excluded from this analysis. Filled circles: full width at half maximum surface brightness. Empty circles: full width at 1\% of the peak surface brightness. $1\sigma$ errors are shown. \label{fig:rp:fwhm}}
\end{center}
\end{figure}

\begin{table}
\centering\footnotesize
\caption{Excess counts over the \textit{Chandra} PSF in the cone and cross-cone wedges 
(excluding the central 1.5\arcsec~(0.834 kpc) nuclear region) for select energy bands. \label{tbl:cts}}
\scriptsize
\begin{tabular}{crlrrlrrl}
\hline\hline
\multirow{2}{*}{Energy} & \multicolumn{2}{c}{8\arcsec circular} & & \multicolumn{2}{c}{S - N} & & \multicolumn{2}{c}{W - E} \\ \cline{2-3} \cline{5-6} \cline{8-9}
(keV) & \multicolumn{2}{c}{Counts (Err)} & & \multicolumn{2}{c}{Counts (Err)} & & \multicolumn{2}{c}{Counts (Err)} \\ \hline
0.3-1.5 & 442.3 & (21.0) & & 293.8 & (17.1) & & 148.5 & (12.2) \\
1.5-3.0 & 238.3 & (15.4) & & 177.3 & (13.3) & & 61.0 & (7.8) \\
3.0-4.0 & 62.0 & (7.9) & & 44.3 & (6.7) & & 17.7 & (4.2) \\
4.0-5.0 & 31.7 & (5.6) & & 25.3 & (5.0) & & 6.4 & (2.5) \\
5.0-6.0 & 23.7 & (4.9) & & 13.3 & (3.7) & & 10.4 & (3.2) \\
6.1-6.5 & 15.2 & (3.9) & & 11.0 & (3.3) & & 4.2 & (2.0) \\ \hline
0.3-7.0 & 812.0 & (28.5) & & 570.2 & (23.9) & & 241.8 & (15.6) \\ \hline
\end{tabular}
\end{table}


The resolution achieved by \textit{Chandra} is unmatched in the X-rays and provides a unique opportunity to study the detailed morphological characteristics of NGC 7212 on subarcsecond scales (1\arcsec~ = 556 pc). Using the CIAO image analysis tools available in SAOImage ds9\footnote{http://ds9.si.edu}, we investigated the spatial characteristics of NGC 7212 by slicing the X-ray emission into six energy bins and generating corresponding images and radial profiles, following the methodology in \citet{Fab18II}. 

The images were built from $1/8$ subpixel data and adaptively smoothed using \textit{dmimgadapt} in the ds9 CIAO package\footnote{http://cxc.harvard.edu/ciao/gallery/smooth.html} for a $0.5-15$ pixel scale with 5 counts under the kernel for 30 iterations (Figure \ref{fig:img:E}). The smoothing parameters were selected to optimize the details of the extended diffuse emission. 
Each energy sliced image reveals a bright nucleus with fainter diffuse large-scale emission.
Focusing on the top two panels of Figure \ref{fig:img:E}, there is obvious $\sim$kiloparsec-scale extended emission in the soft energy bands ($< 3$ keV). Likewise, at higher energies, focusing in particular on the $6.1-6.5$ keV regime where we expect strong Fe K fluorescence (bottom, right panel), extended diffuse emission is present, albeit on smaller scales than in the soft X-rays.

From the $1/8$ subpixel data that was used to generate the smoothed images, we extracted radial surface brightness profiles to quantify the significance of the extended emission. Based on an azimuthal projection of the surface brightness radiating outward from the central nucleus, we slice our data into two cone regions opening outward from the central nucleus in the south$-$north (cone) and west$-$east (cross-cone) directions (Figure \ref{fig:img:full}; right). Interestingly, NGC 7212 does not exhibit a strong azimuthal dependence, unlike what has been found for many other extended hard X-ray sources (e.g., NGC 1068, \citealt{Bau15}; ESO 428$-$G014, \citealt{Fab18II}). Thus we define our cone angles by opposing $90 \degree$ angle wedges centered around the cardinal points (e.g., the south$-$north cone is defined by $90 \degree$ angles that are $\pm45 \degree$ around the north and south axes).
The cone opening angles that we assume in this work are a conservative estimate. 

Concentric annuli were generated out to 8.0\arcsec~(4.448 kpc) ~for each energy band in SAOImage ds9, starting with a width of 1.0\arcsec~(0.556 kpc) and increasing as necessary in the outer regions to maintain a minimum of 10 counts. 
These extracted surface brightness profiles were then background subtracted and compared to a set of \textit{Chandra} Point-source Functions (PSF) generated with ChaRT\footnote{http://cxc.harvard.edu/ciao/PSFs/chart2/} and MARX 5.4.0\footnote{https://space.mit.edu/cxc/marx/} following the CIAO PSF simulation thread\footnote{http://cxc.harvard.edu/ciao/threads/psf.html}\footnote{http://cxc.harvard.edu/ciao/threads/marx\_sim/} for the given centroid positions and energy bands (Figure \ref{fig:rp:full}). 
The PSF models were normalized to the source counts in the central 1.0\arcsec~ circular region.
Within the 8.0\arcsec~ radius and excluding the nuclear region (inner 1.0\arcsec~ circle), the full energy band ($0.3-7.0$ keV) contains $570.2 \pm 23.9$ net excess counts above the \textit{Chandra} PSF in the south$-$north cone and $241.8 \pm 15.6$ net excess counts in the west$-$east cross-cone (Figure \ref{fig:rp:full}).

We further explore these excess counts as a function of energy in each cone region (Table \ref{tbl:cts}). Of note, the high energy band where we would expect to see Fe K fluorescence ($6.1-6.5$ keV) contains a significant excess of $15.2 \pm 3.9$ total counts. The radial profiles for the south$-$north and west$-$east regions as a function of energy are shown in Figures \ref{fig:rp:sn} and \ref{fig:rp:we}, respectively. 
In both the north and east quadrants, the surface brightness falls below the PSF in the inner parsec (0.556 kpc). 
We do not expect this to be caused by pileup, as we estimate using PIMMS\footnote{PIMMS v4.10; http://cxc.harvard.edu/toolkit/pimms.jsp} less than 1\% pileup for this source for both the individual and merged observations.
Rather, this feature may be a consequence of CT obscuration or even strong nuclear absorption by a dust lane as suggested by optical observations along the northeast direction. 

Outside of this feature, we find extended emission on kiloparsec scales (up to 8.0\arcsec $\sim$ 4.5 kpc) in the cone and cross-cone directions, although less significant in the cross-cone region. Analyzing the surface brightness on larger azimuthal scales outside of the 8.0\arcsec~circular region becomes challenging due to possible contamination from the interacting group members, especially in the northern cone. 

Following the methodology of \citet{Fab18II}, we compare the extent of the diffuse emission in each energy bin by calculating the FWHM of the radial profiles in log space. This essentially normalizes the brightness in each energy band, minimizing the bias in the measured extent caused by higher signal-to-noise ratios. This is especially true at low energies where there are significantly more counts.
We fit the radial profiles using a spline approximation, or for energies requiring wider bins (and therefore fewer points) we use a gaussian+polynomial curve. The errors (corresponding to $1\sigma$) are derived from a Monte Carlo error analysis driven by the uncertainty associated with the adaptive binning. 

We find that the FWHM decl.reases slightly with increasing energy in the cone direction (as in \citealt{Fab18II}), but there is no significant trend in the FWHM with energy in the cross-cone direction (Figure \ref{fig:rp:fwhm}; filled circles). The average width is consistent between the cone (red) and cross-cone (blue) directions, $2.10\pm0.23$\arcsec~ ($\sim$1.17 kpc) and $2.15\pm0.75$\arcsec~ ($\sim$1.20 kpc), respectively. 

To better understand the extent of the surface brightness, we also calculate the width at 1\% of the peak emission in each energy bin (Figure \ref{fig:rp:fwhm}; open circles). Compared to the FWHM calculations, we find a larger discrepancy between the cone and cross-cone directions. Below $\sim$4 keV, the average width in the cone direction is $6.50\pm0.47$\arcsec ~($\sim$3.66 kpc) compared to $4.41\pm0.59$\arcsec ~($\sim$2.45 kpc) in the cross-cone direction. Above $\sim$4 keV, the 1\% width drops to $3.83\pm0.12$\arcsec ~($\sim$2.13 kpc) between 5.0 and 7.0 keV, and becomes consistent with the extent in the cross-cone direction. 
Similar trends in the surface brightness extent (as a function of energy) have been observed for ESO 428$-$G014 (\citealt{Fab18II}), for which extended emission in both the cone and cross-cone direction are observed, including that the cross-cone extent of ESO 428$-$G014 drops at lower energies compared to the cone direction.


\section{Spectral Analysis}\label{sec:spec}

\begin{figure}
\begin{center}
\resizebox{70mm}{!}{\includegraphics{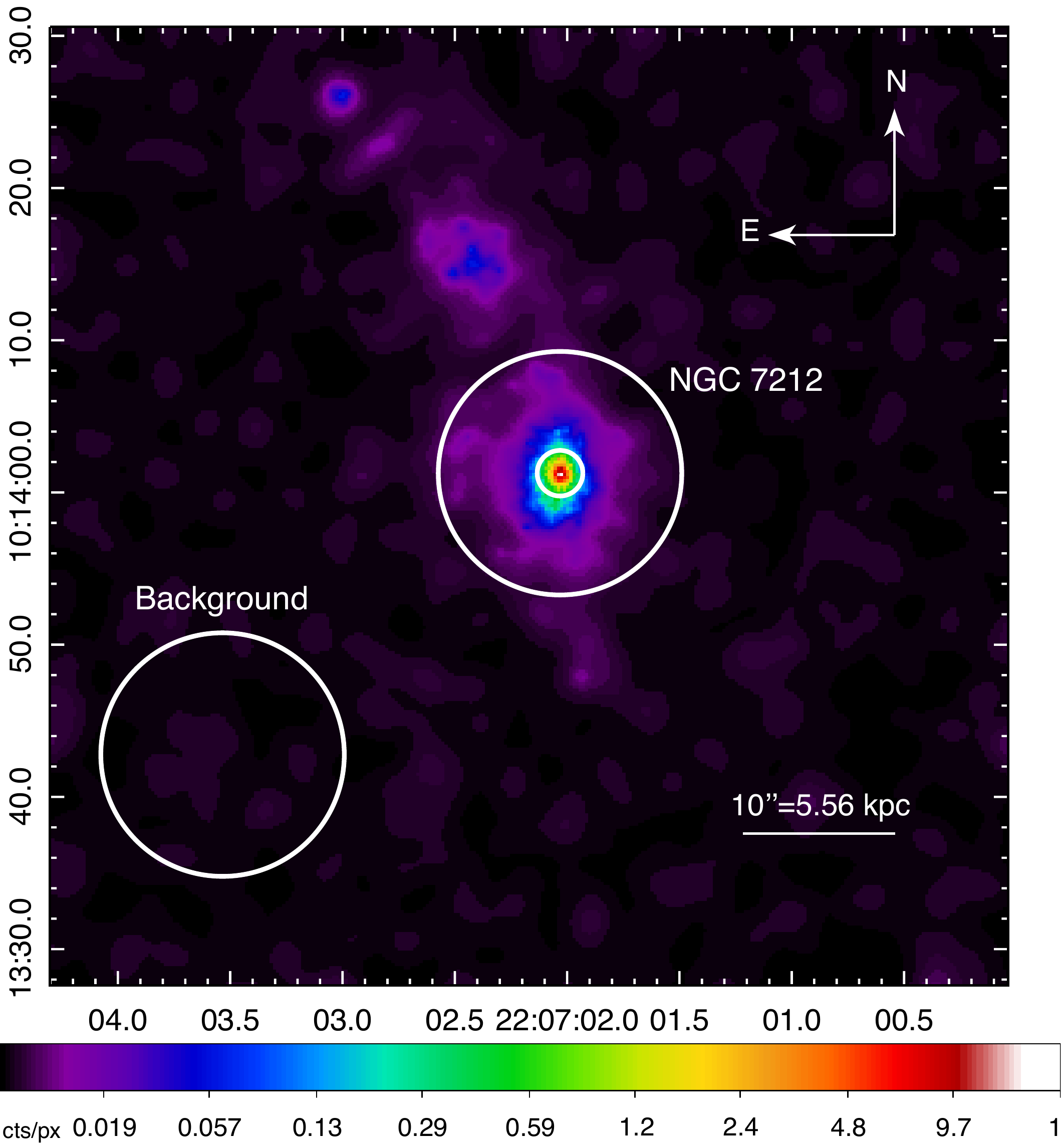}} \\
\caption{Spectral extraction regions overlaid on the merged $0.3-7.0$ keV \textit{Chandra} ACIS image of NGC 7212 (Figure \ref{fig:img:full}; right). The spectra of NGC 7212 is extracted in three regions centered at (J2000) RA = 22:07:02.03 (331$\degree$45$\arcmin$30.44$\arcsec$), decl. = +10:14:01.27 (10$\degree$14$\arcmin$1.27$\arcsec$):  (1) 8.0\arcsec~= 4.448 kpc circular region; (2) 1.5\arcsec~= 0.834 kpc nuclear region; and (3) 1.5\arcsec$-$8.0\arcsec~annulus.
The background region is centered on (J2000) RA = 22:07:03.54 (331$\degree$45$\arcmin$53.00$\arcsec$), decl. = +10:13:42.78 (10$\degree$13$\arcmin$42.78$\arcsec$) with a radius of 10$\arcsec$. \label{fig:img:reg}}
\end{center}
\end{figure}


\begin{figure*}[t!]
\begin{center}
\resizebox{140mm}{!}{\includegraphics{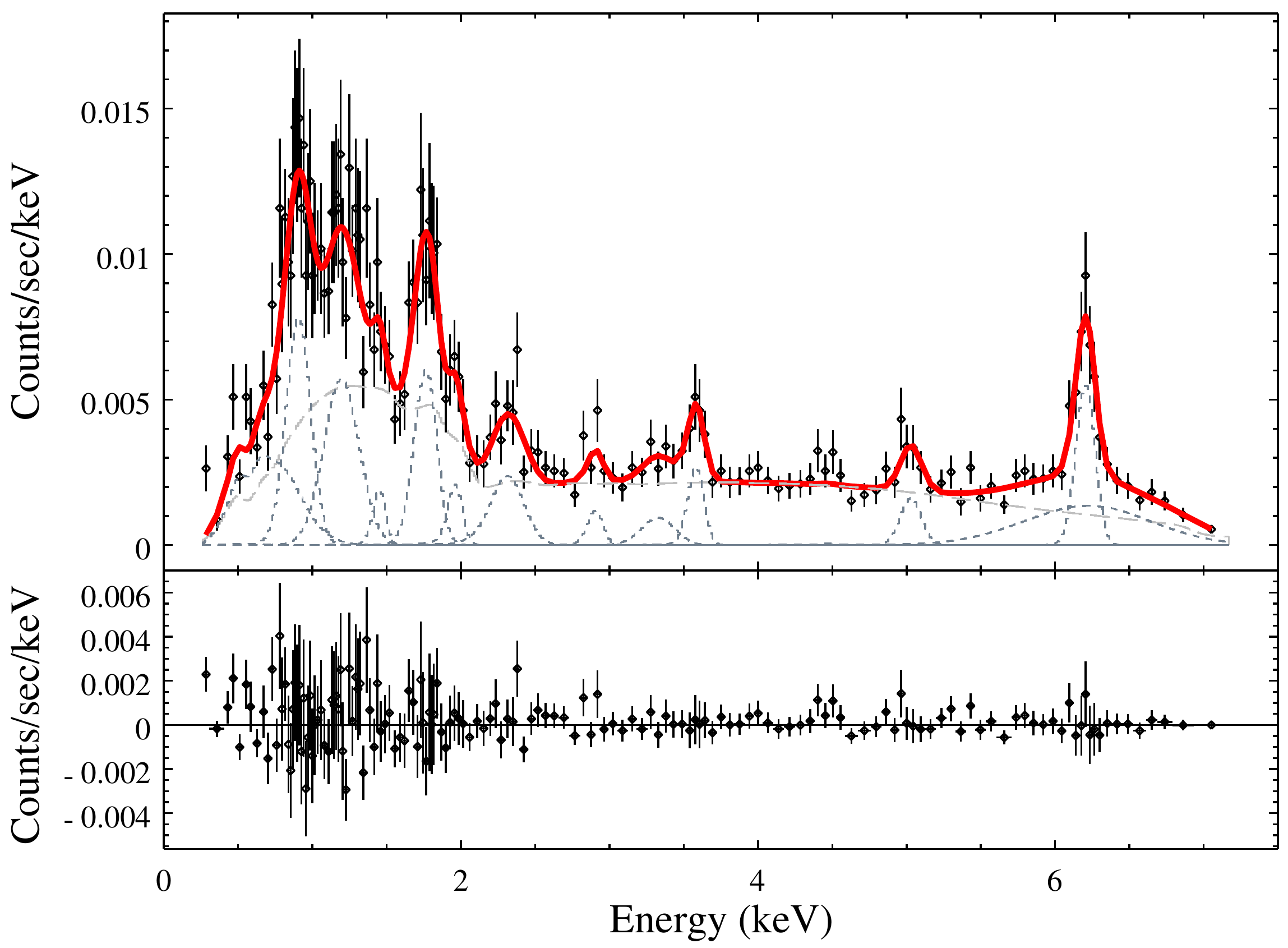}} \\
\caption{ NGC 7212 spectrum extracted from an 8.0\arcsec~(4.448 kpc) circular region centered on the nucleus with best-fit PEXRAV ($\Gamma=1.9$) + \textit{n}-gaussian lines + galactic absorption model (top) and best-fit residuals (bottom). Fit information may be found in Table \ref{tbl:lines}.\label{fig:sp:line}}
\end{center}
\end{figure*}

To characterize the extended X-ray emission, we extracted the spectrum of NGC 7212 for three regions centered on the peak counts at (J2000) RA = 22:07:02.03 (331$\degree$45$\arcmin$30.44$\arcsec$), decl. = +10:14:01.27 (10$\degree$14$\arcmin$1.27$\arcsec$); (1) 8.0\arcsec~= 4.448 kpc circular region, (2) 1.5\arcsec~= 0.834 kpc nuclear region; and (3) 1.5\arcsec$-$8.0\arcsec~annulus (Figure \ref{fig:img:reg}). The background was extracted from a surrounding, off-center, source-free region at (J2000) RA = 22:07:03.54 (331$\degree$45$\arcmin$53.00$\arcsec$), decl. = +10:13:42.78 (10$\degree$13$\arcmin$42.78$\arcsec$) (Figure \ref{fig:img:reg}).
We binned the spectra to have a minimum of 20 counts bin$^{-1}$ in the 8.0\arcsec~circular region and 1.5\arcsec~ nuclear region and a minimum of 10 counts bin$^{-1}$ in the 1.5\arcsec$-$8.0\arcsec~annulus region and fit them to models using Sherpa in the $0.3-7.0$ keV energy band. The $0.3-7.0$ keV spectra extracted from the 8.0\arcsec~circular region is shown in Figure \ref{fig:sp:line}. NGC 7212 has a complex soft excess, strong Fe K$\alpha$ emission, and clear, distinct emission lines between 2 and 6 keV.

\subsection{PEXRAV + Emission Line Models}\label{ssec:PL}

We first fit the hard continuum in each region using a simple reflection PEXRAV model (fold\_E=300; rel\_refl=100; abund,\,Fe\_abund=1; cosIncl=0.45) with power-law photon index $\Gamma=1.9$ (based on the \textit{NuSTAR}/\textit{XMM} best fits, \citealt{Mar18}; and consistent with, e.g., \citealt{Ris00,Lev06,Her15,Ric15,Kos16}), plus a gaussian emission line constrained to an energy range surrounding the Fe K$\alpha$ $6.4$ keV line, with constant galactic absorption ($4.5\times10^{20}$ cm$^{-2}$; \citealt{Lev06}). This simple power-law plus line model is consistent with the model components utilized in previous spectral fits of NGC 7212 (e.g., \textit{ASCA}, \citealt{Ris00}; \textit{XMM}, \citealt{Gua05,Her15}; \textit{CXC}, \citealt{Lev06,Her15}; \textit{NuSTAR}, \citealt{Kos16,Mar18}), for which the median photon index is $\Gamma\sim1.9$ and the median equivalent width of the Fe K$\alpha$ line is $\text{EW}\sim0.8$ keV.

To this simple model we then systematically add unresolved emission lines, allowing the energy and amplitude to vary, while the redshift is kept frozen. 
We use a combination of fit statistics, significance of the emission line fluxes, and visual inspection of the residuals to justify the addition of another line.
The best-fit emission line models are listed in Table \ref{tbl:lines}.

For each region of interest, we find blended emission lines below $\sim$1.5 keV and distinct emission lines above $\sim$1.5 keV with the most significant lines at $\sim$1.8 keV (\ion{Si}{13}) and at $\sim$6.4 keV where we expect to see the Fe K$\alpha$ fluorescent line. There is a degree of uncertainty in the measured energies and amplitudes of the low energy blended lines, but they are consistent with lines (e.g., \ion{O}{7}, \ion{Ne}{9}, \ion{Mg}{11}) observed in other AGN spectra (e.g., \citealt{Kos15,Fab18II,Mak19}) and identified in the NIST Atomic Spectra Database\footnote{http://physics.nist.gov}. 

The $0.3-7.0$ keV spectrum extracted from the 8.0\arcsec~circular region is shown in Figure \ref{fig:sp:line} with the best-fit PEXRAV and line models. 
In the hard X-rays (3.0$-$7.0 keV) and for all regions, we find the characteristic Fe K$\alpha$ emission line. The Fe K$\alpha$ emission is dominated by the neutral emission at $\sim$6.4 keV. The fit benefited from the addition of a broad weak emission line surrounding the neutral Fe line, potentially caused by neutral Fe wings or the presence of blended \ion{Fe}{25} emission. 

We also find strong, significant emission lines in the hard X-rays around 2.9, 3.6, and 5.2 keV (redshift corrected for the distance of NGC 7212). The presence of these lines is not expected for a typical CT AGN (e.g., \citealt{Kos15,Mak19}) and presents an interesting challenge for line classification. Our best identifications for the $\sim$2.9 and $\sim$3.6 keV line are species of calcium fluorescence lines, or varieties of argon (\ion{Ar}{17}, argon K$\alpha$ fluorescence lines). 
The $\sim$5.2 keV emission line that appears to be confined to the inner nuclear region has not yet been identified. It is possible that we are observing the effects of cosmic spallation of the obscuring material such that vanadium K$\alpha$ emission is enhanced (e.g., \citealt{Ski97,Tur10,Gal19}), similar to observations of M51 (\citealt{Xu16}).
It is unlikely that this observed emission is due to the ACIS background, which is fairly well understood at these energies\footnote{http://cxc.harvard.edu/cal/Acis}. 

\begin{table*}
\centering\footnotesize
\caption{Spectral Fitting with PEXRAV + Individual Emission Lines\footnote{Includes galactic absorption of $4.5\times10^{20}$ cm$^{-2}$ (\citealt{Lev06})}}\label{tbl:lines}
\scriptsize
\begin{tabular*}{\linewidth}{@{\extracolsep{\fill}} lcccc}

\hline\hline
 & Counts & Norm. PEXRAV & $\chi_\nu^2$ ($\nu$) &\\
\multicolumn{1}{c}{Region} & (error) & (ph cm$^{-2}$ s$^{-1}$) & Continuum + Lines &\\ \hline

8.0\arcsec~circular region			&	1289 (36)	&	$1.8 \times 10^{-5}$ 		& 0.71 (155) 	&\\
1.5\arcsec~nuclear region			& 	625 (25) 	& 	$1.5 \times 10^{-5}$ 		& 0.67 (218) 	&\\
1.5\arcsec$-$8.0\arcsec~annulus 	& 	664 (26) 	& 	$1.5 \times 10^{-6}$ 		& 0.80 (86) 	&\\ \hline\hline

\multicolumn{1}{c}{Region} & Energy (keV) & Flux ($10^{-6}$ ph cm$^{-2}$ s$^{-1}$) & Significance ($\sigma$) & Identification ($E_{lab}$ keV)\footnote{Energies from NIST (physics.nist.gov); \citealt{Kos15,Mak19}} \\ \hline

8.0\arcsec~circular region & $0.46 ~\substack{+0.11 \\ -0.42}$ &  $42.0 ~\substack{+44.1 \\ -14.9}$ & 	$<3$ 	& \multirow{3}{*}{\shortstack[c]{\ion{N}{7} Ly$\alpha$\footnote{Lines blended in ACIS-S spectrum. These are tentative identifications. \label{fn:blend}} (0.500)\\ \ion{O}{7}\footref{fn:blend} (0.569)}}  \\
1.5\arcsec~nuclear region & $0.49 ~\substack{+0.01 \\ -0.02}$ & $6.7 \pm 2.6$ & 	$<3$ 	&  \\
1.5\arcsec$-$8.0\arcsec~annulus & $0.45 ~\substack{+0.02 \\ -0.03}$ & $10.5 ~\substack{+6.7 \\ -3.8}$ & 	$<3$ 	&  \\ \hline

 	& 	$-$ 	&	$-$ 	&	$-$		&	 \multirow{3}{*}{\ion{Fe}{17}\footref{fn:blend} (0.826)} \\
 	& $0.75 ~\substack{+0.04 \\ -0.05}$ 		& $11.3 ~\substack{+3.1 \\ -2.3}$ & 	$3.7$ 	&  \\
 	& $0.70 ~\substack{+0.05 \\ -0.11}$ 		& $8.1 ~\substack{+1.4 \\ -1.8}$ & 	$4.5$ 	&  \\ \hline
 	& $0.91 ~\substack{+0.01 \\ -0.02}$ 		& $8.7 ~\substack{+1.4 \\ -1.2}$ & 	$6.2$ 	& \multirow{3}{*}{\shortstack[l]{\ion{Ne}{9} (0.905)\\ \ion{Fe}{19} (0.917, 0.922)}} \\
 	& $0.92 ~\substack{+0.02 \\ -0.02}$ 		& $3.1 ~\substack{+1.0 \\ -0.8}$ & 	$3.1$ 	&  \\
 	& $0.93 \pm 0.03$ 					& $2.0 \pm 0.6$ 			& 	$3.3$ 	&  \\ \hline
 	& $1.20 \pm 0.02$ 					& $4.7 \pm 0.7$ 			& 	$6.7$ 	& \multirow{3}{*}{\shortstack[l]{\ion{Fe}{20}\footref{fn:blend} (1.241)\\ \ion{Fe}{24}\footref{fn:blend} (1.129, 1.168)}} \\
 	& $1.14 ~\substack{+0.03 \\ -0.04}$ 		& $1.8 \pm 0.6$ 			& 	$3.0$ 	&  \\
 	& $1.20 \pm 0.03$ 					& $1.7 ~\substack{+0.4 \\ -0.3}$ & 	$4.3$ 	&  \\ \hline
 	& $1.48 \pm 0.03$ 					& $0.4 \pm 0.2$ 			& 	$<3$ 	& \multirow{3}{*}{\shortstack[l]{\ion{Mg}{11} (1.331, 1.352)\\ \ion{Mg}{12} (1.472, 1.745)}} \\
 	& $1.33 \pm 0.02$ 					& $1.5 \pm 0.4$ 			& 	$3.8$ 	&  \\
 	& $1.47 \pm 0.03$, $1.70 \pm 0.02$ 		& $0.41 ~\substack{+0.15 \\ -0.14}$ , $0.52 \pm 0.14$ & 	$3.7$ 	&  \\ \hline
 	& $1.80 \pm 0.01$ 					& $2.3 \pm 0.3$ 			& 	$7.7$ 	& \multirow{3}{*}{\ion{Si}{13} (1.839, 1.865)} \\
 	& $1.81 ~\substack{+0.01 \\ -0.02}$ 		& $1.5 \pm 0.3$ 			& 	$5.0$ 	&  \\
 	& $1.94 \pm 0.04$ 					& $0.57 ~\substack{+0.18 \\ -0.16}$ & 	$3.2$ 	&  \\ \hline
 	& $2.02 \pm 0.02$ 					& $0.5 \pm 0.2$ 			& 	$<3$ 	& \multirow{3}{*}{\ion{Si}{14} (2.005)} \\
 	& $2.01 \pm 0.03$ 					& $0.34 ~\substack{+0.76 \\ -0.16}$ & 	$<3$ 	&  \\
 	& $-$ & $-$ & $-$ &  \\ \hline
 	& $2.37 \pm 0.03$ 					& $2.2 ~\substack{+0.5 \\ -0.4}$ & 	$4.4$ 	& \multirow{3}{*}{\shortstack[l]{S K$\alpha$ (2.308)\\ \ion{S}{15} (2.430)}} \\
 	& $2.40 ~\substack{+0.01 \\ -0.08}$ 		& $0.95 ~\substack{+0.66 \\ -0.22}$ 	& 	$<3$ 	&  \\
 	& $2.32 ~\substack{+0.02 \\ -0.04}$ 		& $0.68 ~\substack{+0.23 \\ -0.20}$ 	& 	$3.0$ 	&  \\ \hline
 	& $2.98 \pm 0.03$ 					& $0.5 \pm 0.2$ 				& 	$<3$ 		& \multirow{3}{*}{\shortstack[l]{Ar K$\alpha$\footnote{Lines in 3-6 keV are rarely observed in AGN. These are tentative line identifications. \label{fn:myst}} (2.958)}} \\
 	& $2.97 \pm 0.03$ 					& $0.30 \pm 0.16$ 				& 	$<3$ 	&  \\
 	& $2.80 \pm 0.05$ 					& $0.63 \pm 0.20$ 				& 	$3.2$ 	&  \\ \hline
 	& $3.68 \pm 0.02$ 					& $1.0 \pm 0.2$ 				& 	$5.0$ 	& \multirow{3}{*}{\shortstack[l]{\ion{Ar}{17}\footref{fn:myst} (3.688)\\ Ca K$\alpha$\footref{fn:myst} (3.691)}} \\
 	& $3.69 ~\substack{+0.02 \\ -0.01}$ 		& $0.78 \pm 0.18$ 				& 	$4.3$ 	&  \\
 	& $3.59 \pm 0.05$ 					& $0.56 ~\substack{+0.17 \\ -0.16}$ 	& 	$3.3$ 	&  \\ \hline
 	& $5.16 ~\substack{+0.02 \\ -0.05}$ 		& $0.9 ~\substack{+0.4 \\ -0.3}$ 	& 	$<3$ 	& \multirow{3}{*}{$-$\footnote{Lines at this energy do not fit into our current understanding of AGN emission. \label{fn:stump}}} \\
 	& $5.17 ~\substack{+0.05 \\ -0.03}$ 		& $1.1 \pm 0.3$ 				& 	$3.7$ 	&  \\
 	& $-$ & $-$ & $-$ &  \\ \hline
 	& $6.37 \pm 0.01$ 					& $5.3 \pm 0.7$ 				& 	$7.6$ 	& \multirow{3}{*}{Fe K$\alpha$ (6.442)} \\
 	& $6.36 \pm 0.01$ 					& $5.0 \pm 0.5$ 				& 	$10.0$ 	&  \\
 	& $6.40 ~\substack{+0.05 \\ -0.07}$ 		& $1.0 \pm 0.3$ 				& 	$3.3$ 	&  \\ \hline
 	& $6.49 ~\substack{+0.08 \\ -0.06}$ 		& $9.1 \pm 1.2$ 				& 	$7.6$ 	& \multirow{3}{*}{\shortstack[l]{Fe K$\alpha$ wing\footref{fn:blend}\\ \ion{Fe}{25}\footref{fn:blend} (6.70)}} \\
 	& $6.64 ~\substack{+0.03 \\ -0.27}$ 		& $8.0 ~\substack{+1.7 \\ -2.4}$ 	& 	$3.0$ 	&  \\
 	& $-$ & $-$ & $-$ &  \\ \hline
\end{tabular*}
\end{table*}

\subsection{Physical Models}\label{ssec:phys}

Beyond understanding what emission lines are found in NGC 7212, it is possible to investigate the physical origin of the X-ray emission using more complex photoionization and thermal spectral models. 
We start building these physical models with the best-fit continuum + Fe K$\alpha$ emission line spectral model for each region, as described in Section \ref{ssec:PL}. To this we add a photoionization and/or thermal model, one component at a time, testing the fit statistics after each addition and estimating the improvement using the \textit{F}-test (Tables \ref{tbl:sp:model}, \ref{tbl:sp:mix}, \ref{tbl:sp:best}; described in subsequent sections). Because the \textit{F}-test has been shown to be unreliable in some cases (e.g., \citealt{Pro02}), we place more emphasis on the fit statistics and observed residuals. We first used purely photoionization and purely thermal models before attempting a mixture of the two.
In situations where only small improvements to the quality of the fit were made by the addition of another component, we examined the residuals of the best fits to identify features that indicate the significance of the improvement (as $\chi^2$ is not sensitive to correlated residuals) to justify incorporating additional complexities.

\begin{figure*}[t!]
\begin{center}
\begin{tabular}{cc}
\resizebox{85mm}{!}{\includegraphics{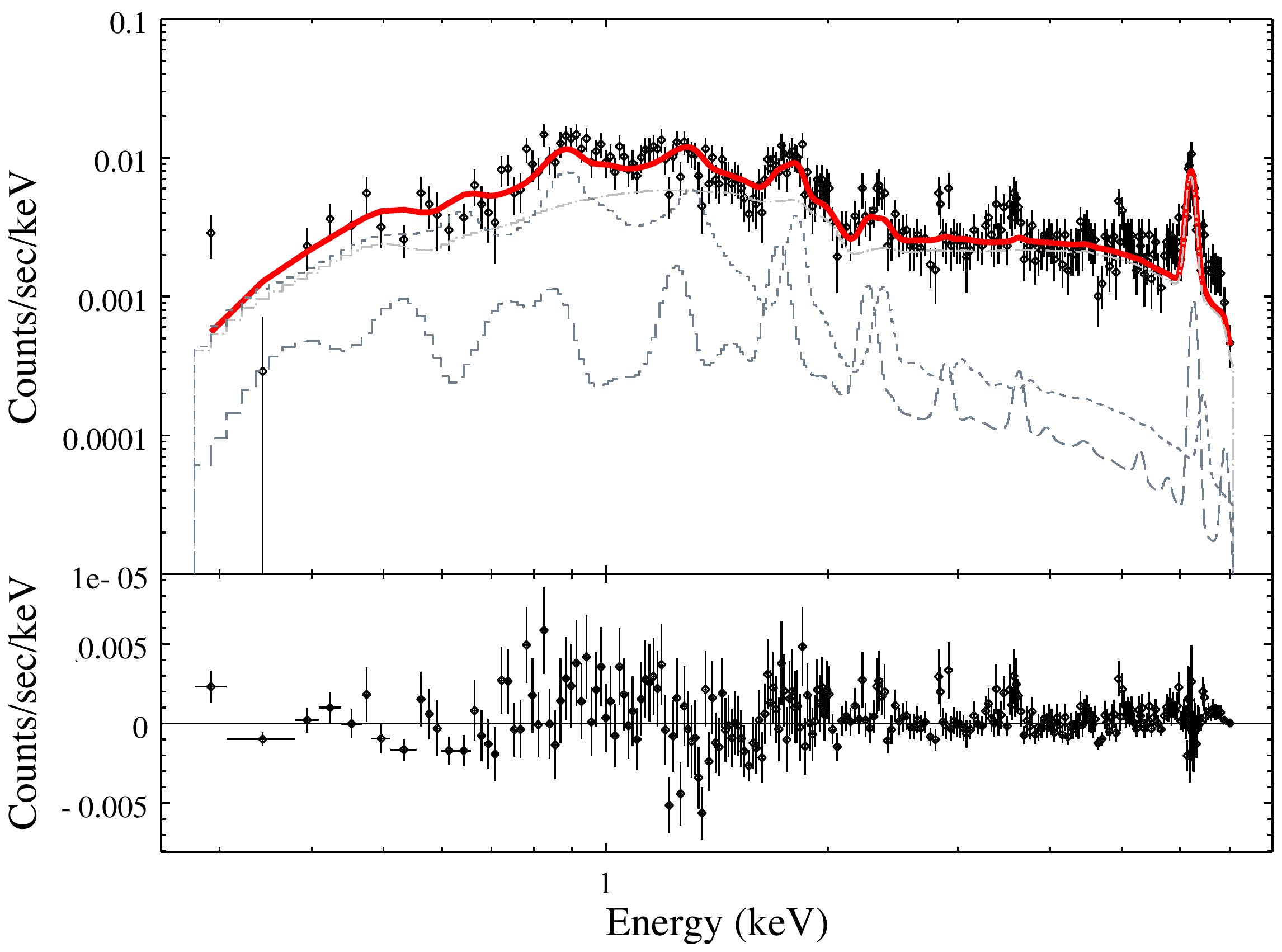}} &  \resizebox{85mm}{!}{\includegraphics{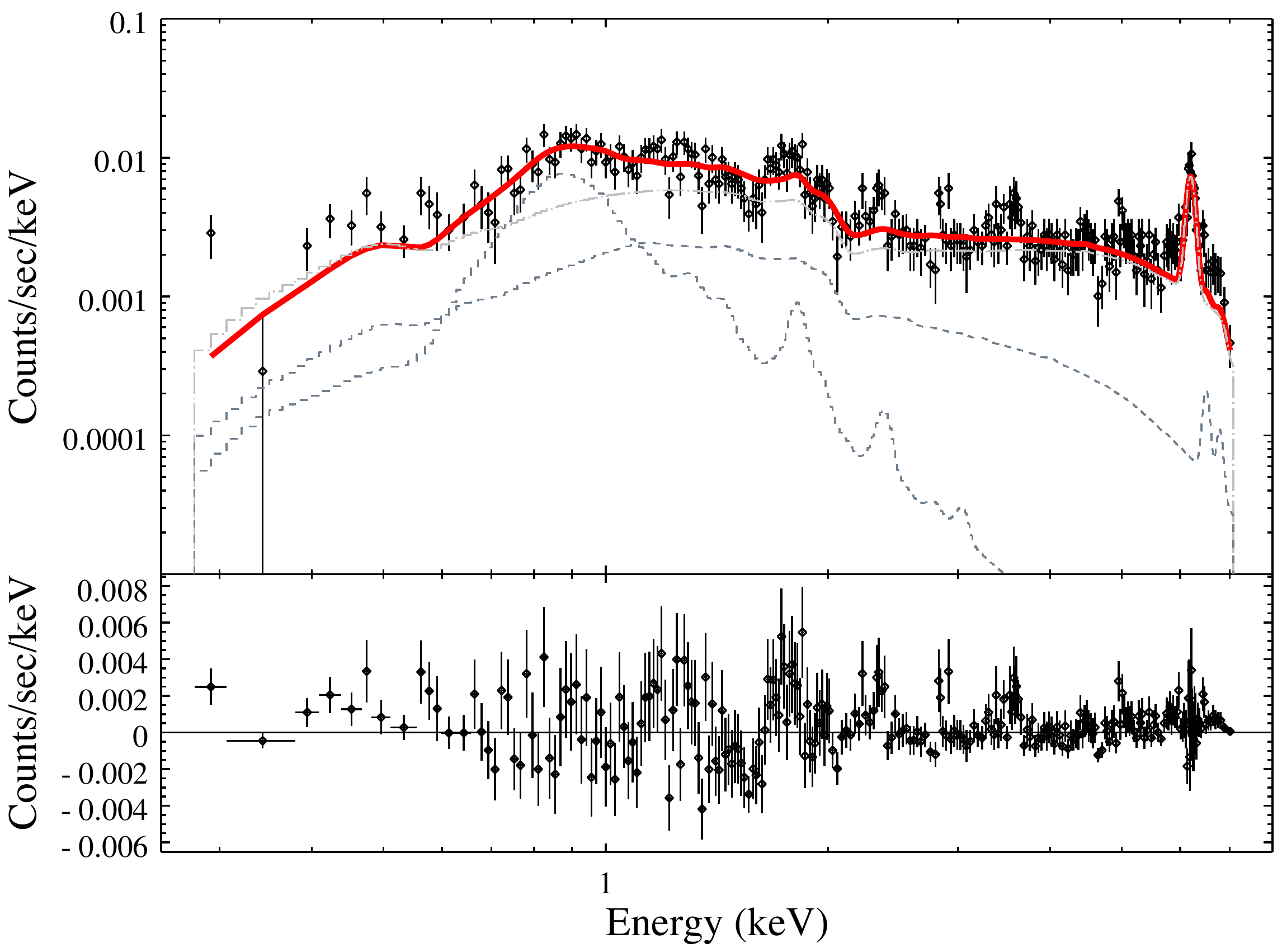}} \\
\resizebox{85mm}{!}{\includegraphics{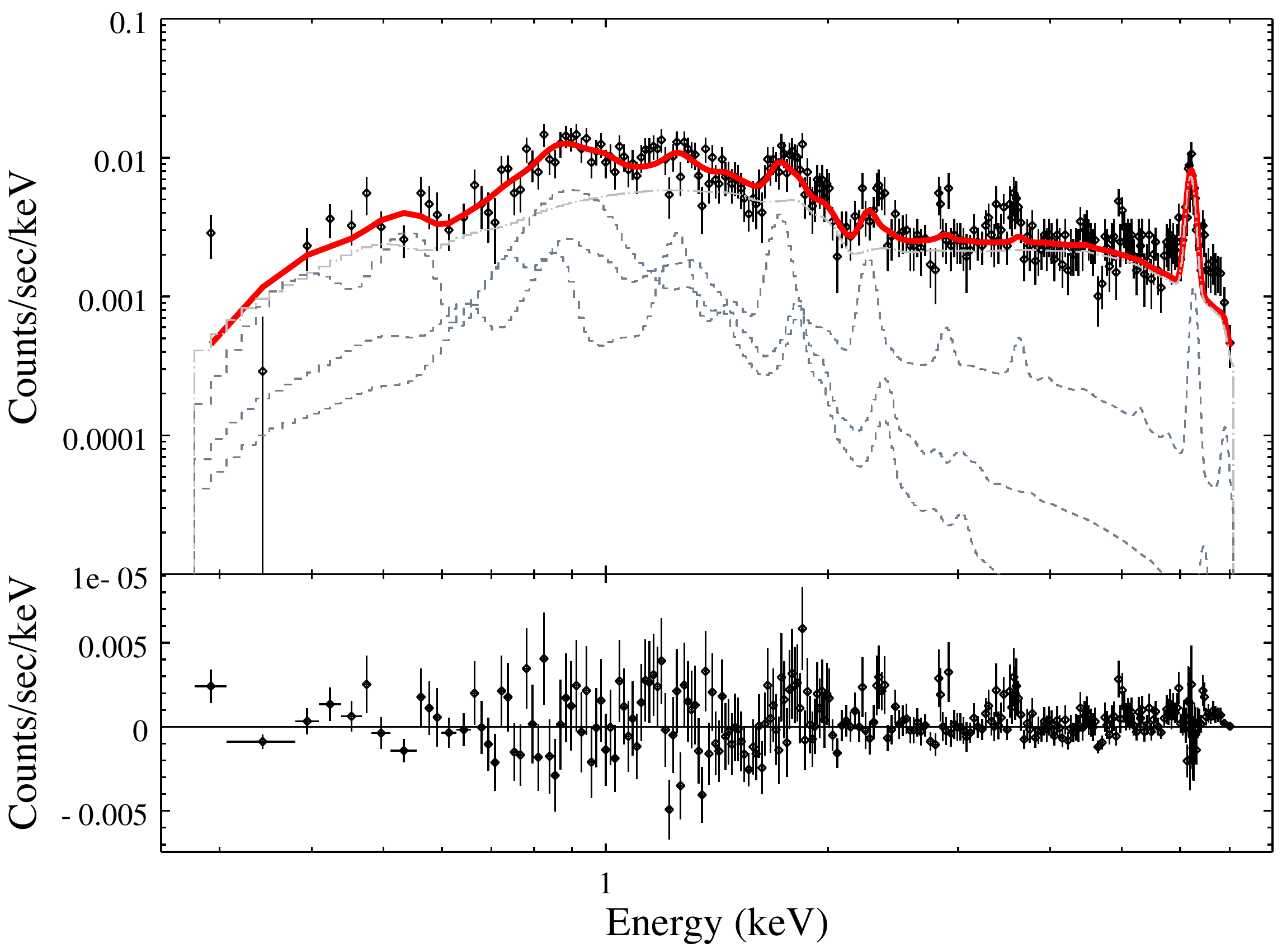}} &  \resizebox{85mm}{!}{\includegraphics{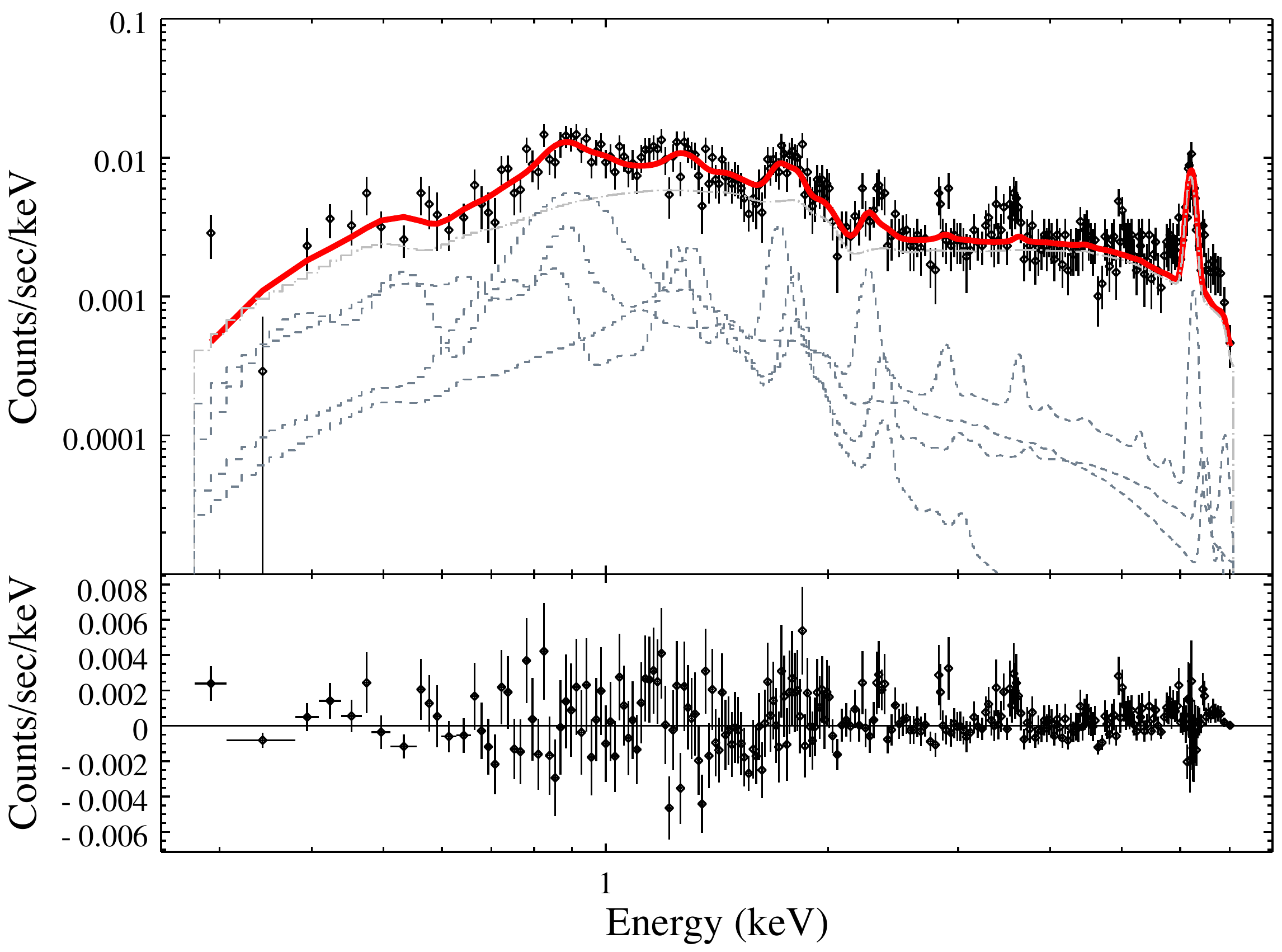}} \\
\end{tabular}
\caption{Top: NGC 7212 spectrum extracted from an 8.0\arcsec~(4.448 kpc) circular region centered on the nucleus fitted with a nuclear PEXRAV component plus (left) two-component photoionization and (right) two-component thermal model. Bottom: the same as above but with a mixture of component models: nuclear PEXRAV component plus a (left) single photoionization plus thermal component model, (right) two-component photoionization plus thermal component model. \label{fig:spec:E}}
\end{center}
\end{figure*}

\subsubsection{N-\textit{component} Photoionization Models}

We start with a photoionization model since one has already been successfully used to describe a lower resolution spectrum of NGC 7212 (\citealt{Bia06}).
The photoionization model that we select is a CLOUDY (\citealt{Fer98}) model of reflection off of a plane parallel slab at an inclination of 45 degrees. 
The intensity and spectral shape of the ionizing radiation are set by the ionization parameter (U) and column density ($\text{N}_\text{H}$), respectively.
Beginning with a single CLOUDY component, we add up to two additional photoionization components. In all three regions, we find a significant improvement increasing from a one- to two-component CLOUDY spectral model and a worse fit adding a third photoionization component (Table \ref{tbl:sp:model}). The best-fit, two-component CLOUDY model parameters are listed in Table \ref{tbl:sp:3fit} and the spectral fit for the $8.0\arcsec$~circular region is shown in Figure \ref{fig:spec:E} (top; left). As shown, this two-component photoionization model cannot fully represent the observed soft X-ray emission or the distinct emission lines between 3 and 6 keV.

\subsubsection{N-\textit{component} Thermal Models}

The thermal components in this investigation are drawn from a solar abundance APEC model (\citealt{Fos12}) with varying temperature (keV) and normalization parameters. We started with a single thermal component and added up to two additional APEC components to our model, although a single component thermal model is not favored (\citealt{Her15}). 
Compared to the pure photoionization models, the fit statistics are worse for the purely thermal components (Table \ref{tbl:sp:model}). 
Furthermore, when comparing the best-fit, two-component CLOUDY model with the best-fit, two-component APEC model residuals, we do not see any improvements in describing the soft X-ray emission. 
The best-fit, two-component APEC model parameters are listed in Table \ref{tbl:sp:3fit} and the spectral fit for the $8.0\arcsec$~circular region is shown in Figure \ref{fig:spec:E} (top; right). 
Of our two thermal components, the components with $kT \sim 0.8$ keV in the $8.0\arcsec$~circular and $1.5\arcsec$~nuclear region are consistent with the APEC model in \citet{Kos16} ($kT \sim 0.8$ keV). 
The addition of a third component significantly improved the fit in the $1.5\arcsec~-8.0\arcsec$~annulus region.

\begin{table}
\footnotesize
\caption{Reduced $\chi^2$, dof, and \textit{F}-Test \textit{P} for Photoionization and Thermal Models}
\label{tbl:sp:model}
\scriptsize
\begin{tabular}{ccclcc}
\hline\hline
 & \multicolumn{2}{c}{Photoionization} &  & \multicolumn{2}{c}{Thermal} \\ \cline{2-3} \cline{5-6} 
 & $\chi_\nu^2 (\nu)$ & \textit{F}-Test \textit{P} &  & $\chi_\nu^2 (\nu)$ & \textit{F}-Test \textit{P} \\ \hline
\multicolumn{2}{l}{(8.0\arcsec~circular region) }  &  &  &  &  \\
$1-$model & 1.20 (257) & ... &  & 1.47 (258) & ... \\
$2-$model & 1.14 (254) & $3.6 \times 10^{-3}$ &  & 1.18 (256) & $8.7 \times 10^{-13}$ \\
$3-$model & 1.15 (251) & $-$ &  & 1.19 (255) & $-$ \\
 &  &  &  &  &  \\
\multicolumn{2}{l}{(1.5\arcsec~nuclear region)}  &  &  &  &  \\
$1-$model & 1.28 (217) & ... &  & 1.34 (123) & ... \\
$2-$model & 1.23 (214) & $4.4 \times 10^{-2}$ &  & 1.21 (121) & $8.1 \times 10^{-7}$ \\
$3-$model & 1.25 (211) & $-$ &  & 1.22 (119) & $-$ \\
 &  &  &  &  &  \\
\multicolumn{2}{l}{(1.5\arcsec$-$8.0\arcsec~annulus)}  &  &  &  &  \\
$1-$model & 1.08 (105) & ... &  & 0.94 (140) & ... \\
$2-$model & 0.88 (102) & $1.5 \times 10^{-4}$ &  & 0.96 (138) & $-$ \\
$3-$model & 0.90 (100) & $-$ &  & 0.77 (136) & $2.9 \times 10^{-7}$ \\ \hline
\end{tabular}
\begin{tablenotes}
\item $^{\text{Note.}}$ n$-$model: number of photoionization or thermal components used in the model.
\end{tablenotes}
\end{table}

\begin{table*}
\centering\footnotesize
\caption{Best-fit Parameters with Two-component Models}
\label{tbl:sp:3fit}
\begin{threeparttable}
\scriptsize
\begin{tabular*}{\textwidth}{@{\extracolsep{\fill}} ll|ll}
\hline\hline
\multicolumn{2}{l|}{Two$-$Photoionization} & \multicolumn{2}{l}{Two$-$Thermal} \\ \hline
(8.0\arcsec~circular region) &  & (8.0\arcsec~circular region) &  \\
$\log(\text{U1}) = 1.35 ~\substack{+0.17 \\ -0.03} $ & $\log(\text{N$_\text{H}$1}) = 23.5 ~\substack{+0.0 \\ -0.2} $ & $k\text{T1} = 0.82 \pm 0.05$ & $\text{EM1} = (8.7 \pm 0.8) \times 10^{-6}$ \\
$\log(\text{U2}) = -1.25 ~\substack{+0.05 \\ -1.75}$ & $\log(\text{N$_\text{H}$2}) = 23.5 ~\substack{+0.0 \\ -0.6}$ & $k\text{T2} = 8.62 ~\substack{+5.21 \\ -2.38}$ & $\text{EM2} = (3.0 ~\substack{+0.2 \\ -0.3}) \times 10^{-5}$ \\
 &  &  &  \\
(1.5\arcsec~nuclear region) &  & (1.5\arcsec~nuclear region) &  \\
$\log(\text{U1}) = 1.30 ~\substack{+0.21 \\ -0.14} $ & $\log(\text{N$_\text{H}$1}) = 23.2 ~\substack{+0.3 \\ -0.4}$ & $k\text{T1} = 0.81 \pm 0.05$ & $\text{EM1} = (6.0 \pm 0.6) \times 10^{-6}$\\
$\log(\text{U2}) = -1.99 ~\substack{+0.50 \\ -1.01} $ & $\log(\text{N$_\text{H}$2}) = 23.3 ~\substack{+0.2 \\ -0.4}$ & $k\text{T2} = 8.62 ~\substack{+20.2 \\ -2.9}$ & $\text{EM2} = (1.7 ~\substack{+0.4 \\ -0.2}) \times 10^{-5}$ \\
 &  &  &  \\
(1.5\arcsec$-$8.0\arcsec~annulus) &  & (1.5\arcsec$-$8.0\arcsec~annulus) &  \\
$\log(\text{U1}) = 1.42 ~\substack{+0.18 \\ -0.10}$ & $\log(\text{N$_\text{H}$1})  = 19.5 ~\substack{+1.2 \\ -0.5}$ & $k\text{T1} = 4.92 ~\substack{+1.32 \\ -0.88}$ & $\text{EM1} = (2.0 \pm 0.2) \times 10^{-5}$  \\
$\log(\text{U2}) = -1.0 ~\substack{+0.03 \\ -0.24}$ & $\log(\text{N$_\text{H}$2}) = 22.7 ~\substack{+0.3 \\ -0.5}$ & $k\text{T2} = 5.23 ~\text{unconstrained}$ & $\text{EM2} = 2.3 \times 10^{-8}$ \text{unconstrained} \\ \hline
\end{tabular*}
\begin{tablenotes}
\item[Note.] U is the ionization parameter of each component, N$_\text{H}$ is the column density (cm$^{-2}$), $k$T is the temperature (keV), and EM is the normalization of the APEC model (cm$^{-5}$); $\text{EM}=\frac{10^{-14}}{4\pi [D_A(1+z)]^2}\int n_e n_H dV$, where $D_A$ is the angular diameter distance to the source (cm) and $n_e$, $n_H$ are the electron and Hydrogen densities (cm$^{-3}$).
\end{tablenotes}
\end{threeparttable}
\end{table*}

\subsubsection{Mixed Photoionization and Thermal Models}

Since fits using individual photoionization and thermal models failed to adequately represent the complex observed emission, we fit the spectrum with a variety of mixed model combinations, up to three each. The fit statistics and \textit{F}-test results for the selected model mixtures in each of the three regions are shown in Table \ref{tbl:sp:mix}. 

The best model combinations are region dependent. 
In the $1.5\arcsec$~nuclear region, the $2 + 1$ (two-photoionization, single thermal) model offered the best fit. Similarly this fit worked well in the $8.0\arcsec$~circular region. For the $8.0\arcsec$~circular region, including an additional thermal component ($2 + 2$) improved the spectral fit slightly. Adding additional thermal components ($1 + 2$) improved the spectral fit for the $1.5\arcsec$ nuclear region compared to the single photoionization and thermal model, both in the observed residuals and the \textit{F}-test calculation. 

The best-fit parameters for our mixed models are listed in Table \ref{tbl:sp:mixfit} and the $2 + 1$ and $2 + 2$ spectral fits for the $8.0\arcsec$~circular region are shown in Figure \ref{fig:spec:E} at the bottom right and left, respectively. 
In the extended emission region ($1.5\arcsec-8.0\arcsec$~annulus), all of the mixed model spectral fits have $\chi^2 \lesssim 1$ and thus are acceptable solutions. 
In each region, however, none of the mixed models are able to completely fit the distinct emission lines between 3 and 6 keV, in which we see notable residual features for every mixed model combination.

\begin{table*}
\centering\footnotesize
\caption{Reduced $\chi^2$, dof, and F-Test P for Mixed Photoionization and Thermal Models}
\label{tbl:sp:mix}
\scriptsize
\begin{tabular*}{\textwidth}{@{\extracolsep{\fill}}c|cc|cc|cc}
\hline\hline
 \multicolumn{1}{c}{} & \multicolumn{2}{c}{8.0\arcsec~circular region} & \multicolumn{2}{c}{1.5\arcsec~nuclear region} & \multicolumn{2}{c}{1.5\arcsec$-$8.0\arcsec~annulus} \\
N-Photo + N-Thermal & $\chi_\nu^2 (\nu)$ & F-Test P & $\chi_\nu^2 (\nu)$ & F-Test P & $\chi_\nu^2 (\nu)$ & F-Test P \\ \hline
$1 + 1$ & 1.18 (255) & ... & 1.23 (215) & ... & 0.77 (137) & ... \\
$1 + 2$ & 1.12 (253) & $2.4 \times 10^{-3}$ & 1.23 (213) & $-$ & 0.70 (135) & $2.1 \times 10^{-3}$ \\
$1 + 3$ & 1.09 (251) & $1.7 \times 10^{-2}$ & 1.24 (211) & $-$ & 0.72 (133) & $-$ \\
 &  &  &  &  &  &  \\
$2 + 1$ & 1.08 (252) & $8.5 \times 10^{-5}$ & 1.197 (212) & $0.18$ & 0.72 (134) & $2.7 \times 10^{-2}$ \\
$2 + 2$ & 1.07 (250) & $0.27$ & 1.199 (210) & $-$ & 0.73 (132) & $-$ \\
$2 + 3$ & 1.08 (248) & $-$ & 1.24 (208) & $-$ & 0.74 (130) & 0.054 \\
 &  &  &  &  &  &  \\
$3 + 1$ & 1.08 (249) & $0.98$ & 1.23 (209) & $-$ & 0.75 (131) & $-$ \\
$3 + 2$ & 1.08 (247) & $-$ & 1.26 (207) & $-$ & 0.82 (129) & $-$ \\
$3 + 3$ & 1.11 (245) & $-$ & 1.27 (205) & $-$ & 0.75 (127) & $-$ \\ \hline
\end{tabular*}
\end{table*}

\begin{table*}
\centering\footnotesize
\caption{Summary of Reduced $\chi^2$, dof, and F-Test P for Best-fit N-Photoionization + N-Thermal Models}
\label{tbl:sp:best}
\scriptsize
\begin{tabular*}{\textwidth}{@{\extracolsep{\fill}} ccc|ccc|ccc}
\hline\hline
\multicolumn{3}{c}{8.0\arcsec~circular region} &  \multicolumn{3}{c}{1.5\arcsec~nuclear region} &  \multicolumn{3}{c}{1.5\arcsec$-$8.0\arcsec~annulus} \\
P $+$ T& $\chi_\nu^2 (\nu)$ & F-Test P & P $+$ T & $\chi_\nu^2 (\nu)$ & F-Test P & P $+$ T & $\chi_\nu^2 (\nu)$ & F-Test P \\ \hline
$0 + 2$ & 1.18 (256) & ... & $0 + 2$ & 1.21 (269) & ... & $0 + 3$ & 0.77 (136) & ... \\
$2 + 0$ & 1.14 (254) & $1.3 \times 10^{-2}$ & $2 + 0$ & 1.23 (214) & $-$ & $1 + 2$ & 0.70 (135) & $3.4 \times 10^{-4}$ \\
$2 + 1$ & 1.08 (252) & $1.1 \times 10^{-3}$ & $2 + 1$ & 1.197 (212) & $5.6 \times 10^{-2}$ & $2 + 0$ & 0.88 (102) & $-$ \\
$2 + 2$ & 1.07 (250) & 0.31 & & & & & & \\ \hline
\end{tabular*}
\end{table*}

\begin{table*}
\centering\footnotesize
\caption{Best-fit Parameters with Mixed Models}
\label{tbl:sp:mixfit}
\begin{threeparttable}
\scriptsize
\begin{tabular*}{\textwidth}{@{\extracolsep{\fill}} ll|ll }
\hline\hline
\multicolumn{2}{l|}{8.0\arcsec~circular region: Two$-$Photoionization + Thermal} & \multicolumn{2}{l}{8.0\arcsec~circular region: Two$-$Photoionization + Two$-$Thermal} \\ \hline
$\log(\text{U1}) = 1.25 ~\substack{+0.07 \\ -0.05} $ & $\log(\text{N$_\text{H}$1}) = 21.7 ~\substack{+1.8 \\ -2.3} $ & $\log(\text{U1}) = 1.25 ~\substack{+0.09 \\ -0.22} $ & $\log(\text{N$_\text{H}$1}) = 23.5 ~\substack{+0.0 \\ -0.4}$ \\
$\log(\text{U2}) = -1.04 ~\substack{+0.08 \\ -0.17} $ & $\log(\text{N$_\text{H}$2}) = 23.3 ~\substack{+0.2 \\ -0.3} $ & $\log(\text{U2}) = -1.23 ~\substack{+0.23 \\ -0.26} $ & $\log(\text{N$_\text{H}$2}) = 23.4 ~\substack{+0.1 \\ -0.3}$ \\
$k\text{T1} = 0.88 ~\substack{+0.07 \\ -0.05} $ & $\text{EM1} = (6.6 ~\substack{+0.9 \\ -0.8}) \times 10^{-6}$ & $k\text{T1} = 0.89 ~\substack{+0.07 \\ -0.06}$ & $\text{EM1} = (6.4 \pm 0.9) \times 10^{-6}$ \\
& & $k\text{T2} = 5.44 ~\substack{\text{+ ~---~}\\-3.22}$ & $\text{EM2} = (7.4 ~\substack{+2.8 \\ -2.7}) \times 10^{-6}$ \\
\hline
\multicolumn{2}{l|}{1.5\arcsec~nuclear region: Two$-$Photoionization + Thermal} & \multicolumn{2}{l}{1.5\arcsec$-$8.0\arcsec~annulus: Photoionization + Two$-$Thermal} \\ \hline
$\log(\text{U1}) = 1.25 ~\substack{+0.09 \\ -0.19} $ & $\log(\text{N$_\text{H}$1}) = 23.5 ~\substack{+0.0 \\ -0.5} $ & $\log(\text{U1}) = -0.75 ~\substack{+0.22 \\ -0.26} $ & $\log(\text{N$_\text{H}$1}) = 22.9 ~\substack{+0.5 \\ -0.6}$ \\
$\log(\text{U2}) = -1.50 ~\substack{+0.25 \\ -0.47} $ & $\log(\text{N$_\text{H}$2}) = 23.3 ~\substack{+0.2 \\ -0.3} $ & $k\text{T1} = 0.86 ~\substack{+0.09 \\ -0.04}$ & $\text{EM1} = (2.8 ~\substack{+0.5 \\ -0.4}) \times 10^{-6}$\\
$k\text{T1} = 0.96 ~\substack{+0.09 \\ -0.11}$ & $\text{EM1} = (3.9 \pm 0.7) \times 10^{-6}$ & $k\text{T2} = 6.84 ~\substack{+7.92 \\ -2.45}$ & $\text{EM2} = (1.1 \pm 0.1) \times 10^{-5}$ \\ \hline
\end{tabular*}
\begin{tablenotes}
\item[Note.] U is the ionization parameter of each component, N$_\text{H}$ is the column density (cm$^{-2}$), $k$T is the temperature (keV), and EM is the normalization of the APEC model (cm$^{-5}$).
\end{tablenotes}
\end{threeparttable}
\end{table*}

\section{Discussion}\label{sec:dis}

Recently, deep, high-resolution observations of CT AGN have uncovered extended hard X-ray emission on $\sim$kpc scales, a challenge to the long-held belief that hard X-ray emission was limited to the central few parsecs of the nucleus (e.g., \citealt{Fab17,Mak17}).
The long-held assumption was that the origin of this emission is the reflection of energetic photons from the corona off obscuring material close to the nucleus (e.g., torus with luminosity-dependent distance of $0.1-1$ pc; \citealt{Net15}). This picture is in line with the ``standard'' AGN model (\citealt{Urr95}), in which the geometrical orientation of an AGN, with respect to the observer, produces its observed multiwavelength properties. The presence of kiloparsec-scale extended emission raises questions about the nature, origin, and locations of the obscuring material.

High-resolution imaging observations of these sources allow us to better understand the limitations of the standard model and provide clues as to how the AGN interacts with and impacts its host galaxy through feedback processes.
This work has closely examined NGC 7212, a CT AGN with extended X-ray emission, to better understand the spectral and spatial characteristics of this AGN class and provide constraints on the connection between AGN and their host galaxies. A discussion of our results is presented in the following subsections.

\subsection{Luminosities}

From our spectral fits we have calculated the $0.3 - 7.0$ keV luminosity for each region of interest (8.0\arcsec~circular region, 1.5\arcsec~nuclear region, and a 1.5\arcsec$-$8.0\arcsec~annulus; Figure \ref{fig:img:full}). We find that the 8\arcsec~ circular region, containing the majority of the galactic emission, has a luminosity of $L_{0.3-7.0}=7.36\times10^{41}$ erg s$^{-1}$. Breaking this down further, we find the inner 1.5\arcsec~ nuclear region, which contains the majority of the CT AGN emission, has a luminosity of $L_{0.3-7.0}=6.07\times10^{41}$ erg s$^{-1}$. 
For comparison with published results, we calculate the luminosity for $2-10$ keV in the 8\arcsec~ region using PIMMS\footnote{http://cxc.harvard.edu/toolkit/pimms.jsp} and find an observed luminosity of $L_{2-10}=4.80\times10^{41}$ erg s$^{-1}$ ($\log L_{2-10}=41.68$). Correcting for absorption, we calculate an intrinsic luminosity of $L_{2-10}=8.03\times10^{42}$ erg s$^{-1}$ ($\log L_{2-10}=42.90$). This is consistent with what has been found for NGC 7212 in previous observations (e.g., observed luminosities: $\log L_{2-10}=41.95$, \citealt{Kos16}; intrinsic, unabsorbed luminosities: $\log L_{2-10}=42.6$, \citealt{Her15}; $\log L_{2-10}=43.07$, \citealt{Kos16}; $\log L_{2-10}=42.8$, \citealt{Mul18}; $\log L_{2-10}=43.21$, \citealt{Mar18}).

\subsection{Spectral Emission Lines}\label{ss:line}

The depth of our observation of NGC 7212 allows us to analyze three separate regions (8.0\arcsec~circular region, 1.5\arcsec~nuclear region, and a 1.5\arcsec$-$8.0\arcsec~annulus) with enough statistical significance that we can compare the properties of the nuclear region to the region of extended emission. Each region is fit using a reflected power-law continuum with $\Gamma = 1.9$, consistent with the work of, e.g., \citet{Kos16} and \citet{Mar18}, and additional gaussian emission lines, including a strong Fe K$\alpha$ component. Not surprisingly, the normalization of the power-law component in the circumnuclear region is almost an order of magnitude larger than for the diffuse, extended emission in the outer region.

While we assume a photon index for the continuum in our emission line models of $\Gamma = 1.9$, we did compare the best-fit index for the nuclear and extended regions in the hard X-rays. We find that the extended region has a steeper photon index compared to that of the nuclear region ($2.56\pm0.06$ compared to $2.10\pm0.02$), similar to what is found in ESO 428$-$G014 (\citealt{Fab17}).

At energies below $\sim$1.2 keV, the predicted emission lines are complex and blended, but are consistent with what is observed in other nearby galaxies (e.g., \citealt{Kos15,Fab18II,Mak19}). Similarly, the presence of strong \ion{Mg}{11} (1.331, 1.352 keV) and \ion{Si}{8} (1.839, 1.865 keV) are typical for CT AGN. 

In the hard X-rays, the most significant emission line that we see can be attributed to the Fe K$\alpha$ emission line near $6.4$ keV. 
This line is believed to originate from within the AGN torus region due to the excitation of the obscuring material. Thus it comes as no surprise that this emission line is strong in the nuclear region (inner 1.5\arcsec $\sim$0.8 kpc) where the CT AGN is located.
However, we also find significant Fe K$\alpha$ emission in the annulus in excess of the \textit{Chandra} PSF model and extending to $\sim$3.7 kpc scales at $\sim20\%$ of the relative strength of the nuclear region. 

In the simplest picture where AGN are classified as a function of line-of-sight orientation (e.g., \citealt{Urr95}), extended Fe K$\alpha$ X-ray emission is unexpected for a CT AGN since the obscuring material (e.g., torus) is expected to act as a screen in the inner $0.1-10$ parsec, limiting this emission to the nuclear region (e.g., \citealt{Net15}). However, recent observations of nearby CT AGN find significant Fe K$\alpha$ emission outside of the central nucleus (e.g., Circinus: extended $\sim$100 pc, \citealt{Mar13,Are14}; NGC 1068: 30\% Fe K$\alpha$ emission observed at $>$140 pc, \citealt{Bau15}; ESO 428$-$G014: extended $\sim$3.7 kpc, \citealt{Fab17,Fab18II,Fab18III}).

The hard energy band of NGC 7212 is more complex than expected for a CT AGN, especially in the 1.5\arcsec~nuclear region, where the AGN emission dominates. We discuss the presence of strong, significant lines between $\sim3$ and 6 keV and their tentative identifications below.

\subsubsection{Unique Features}

We find three emission features not, to our knowledge, previously reported in AGN X-ray spectra. There is significant line emission near $2.9$ keV in the outer annuli ($1.5-8.0\arcsec$; Table \ref{tbl:lines}) comparable in strength to the Fe K$\alpha$ line that may be an argon fluorescence line ($E_{lab}=2.958$ keV). In order to achieve this kind of relative abundance in a photoionized plasma, the annulus region would need to contain a steeper ionizing spectrum compared to the nuclear region (which is consistent with our observations of the relative photon index; see Section \ref{ss:line}). However, since the AGN spectrum is filtered through the torus, it is challenging to know for certain what shape illuminates the ambient gas.
This line may also be present in the central nuclear region, but only at $1.9\sigma$ significance and so is unlikely to originate from the CT AGN directly. 

In all three regions, we find a significant emission line near $3.6$ keV (Table \ref{tbl:lines}). At this energy, possible identifications include \ion{Ar}{17} ($E_{lab}=3.688$ keV) or a calcium K$\alpha$ fluorescence line ($E_{lab}=3.691$ keV). At both $2.9$ and $3.6$ keV we cannot make anything more than tentative identifications since these lines are not characteristic of AGN. Furthermore, we do not currently have a good physical explanation for only finding neutral calcium.

We also find a distinct emission line near $5.17$ keV that is limited to the central 1.5\arcsec~nuclear region. 
We currently have no definitive identification for the emission line at $5.17$ keV, despite a thorough search of literature and the NIST Atomic Spectra Database\footnote{http://physics.nist.gov}. It is unlikely to be an artifact of the ACIS-S background, which is fairly well known at this energy\footnote{http://cxc.harvard.edu/cal/Acis}. Likewise, it is unlikely to be due to contamination from a nearby or background galaxy, as its closest neighbor is located 9.8 kpc away (e.g., \citealt{Kos16}), and the probability of overlap with a background galaxy (e.g., located at $z=0.24$ where Fe K$\alpha$ could explain this observed emission line) is low.
vanadium He-like emission at $5.18$ keV from cosmic spallation (energetic particles bombarding optically thick material leading to the formation of elements) could explain this observed emission (\citealt{Gal19}). However, vanadium K$\alpha$ is typically weak compared to other spallation lines that we do not see (e.g., titanium, chromium; \citealt{Gal19}). 
The importance of spallation in AGN is debatable (e.g., \citealt{Ski97,Tur10,Gal19}), as is the origin of the energetic particles (e.g., accretion disks, disk winds; \citealt{Tur10}). 
The development of high-resolution spectroscopic instruments, e.g, calorimeters, is increasingly important for detecting spallation and identifying emission at unusual energies, especially in the hard X-rays.

\subsubsection{Physical Models}

As discussed in Section \ref{ssec:phys}, we find that a single photoionization or thermal component does not fully represent the soft X-ray emission from NGC 7212 and a more complex mixture of the two is needed. 

For the $1.5\arcsec$~nuclear region ($1.5\arcsec$ = 834 pc), the best fit to the spectrum is a combination of the two photoionization models and a single thermal model with $k\text{T}=0.96~\substack{+0.09 \\ -0.11}$ keV, and the two ionization parameters $\log \text{U}_1=-1.50~\substack{+0.25 \\ -0.47}$ and $\log \text{U}_2=1.25~\substack{+0.09 \\ -0.19}$. 
In comparison, the two-component thermal model finds $k\text{T}=0.81\pm0.05$ keV, and $k\text{T}=8.62~\substack{+20.2 \\ -2.9}$ keV. The two-component photoionization model finds $\log \text{U}=-1.99~\substack{+0.5\\ -1.01}$, and $\log \text{U}=1.30~\substack{+0.21 \\ -0.14}$ (Table \ref{tbl:sp:3fit}). 
Since many of the fit statistics in this region are very similar, we can conservatively say that the best spectral fit is given by, at minimum, one thermal and one photoionization component (but two photoionization models are preferred).

In the $1.5\arcsec-8.0\arcsec$~annulus region, the best-fit physical model is a combination of a single photoionization model with $\log \text{U}=-0.75~\substack{+0.22 \\ -0.26}$, and two thermal models with $k\text{T}=0.86~\substack{+0.09 \\ -0.04}$ keV, and $k\text{T}=6.84~\substack{+7.92 \\ -2.45}$ keV (Table \ref{tbl:sp:mixfit}). The three-component thermal model finds $k\text{T}=0.85~\substack{+0.05 \\ -0.08}$ keV, $k\text{T}=6.85~\substack{+18.7 \\ -2.35}$ keV, and $kT \sim $64 keV, which is essentially pure bremsstrahlung. The two-component photoionization model finds $\log \text{U}=-1.00~\substack{+0.03 \\ -0.24}$, and $\log \text{U}=1.42~\substack{+0.18 \\ -0.10}$.
Similar to the $1.5\arcsec$~nuclear region, a three-component model is preferred in the annulus with, at minimum, one photoionization and one thermal model, plus one additional thermal or photoionization model. 

Looking at the circumnuclear region ($8.0\arcsec$~circular region), we find that the best fit is the combination of two-photoionization and two-thermal models, although it is also well fit by a two-photoionization-one-thermal mixture. The two-two mixture is preferred since the residuals of the best fit are visually less correlated (Figure \ref{fig:spec:E}). This two-two model has the parameters: $k\text{T}=0.89~\substack{+0.07 \\ -0.06}$ keV, $k\text{T}=5.44~\substack{+ ~\text{---}~ \\ -3.22}$ keV, $\text{U}=-1.23~\substack{+0.23 \\ -0.26}$, and $\log \text{U}=1.25~\substack{+0.09 \\ -0.22}$. These are consistent with the best-fit thermal and ionization parameters in both the nuclear and annular regions. Furthermore, the thermal component fit in each region, $kT \sim $0.89 keV, is consistent with previous observations of NGC 7212 (e.g., \citealt{Kos16}; $kT \sim $0.8 keV). The densities that we derive in these mixed models are also consistent with typical densities in the interstellar medium (Tables \ref{tbl:sp:3fit}, \ref{tbl:sp:mixfit}). 

While NGC 7212 requires more than a single model component, the best-fit model mixture in each region was less complex than seen in other sources with observed extended X-ray emission (e.g., ESO 428$-$G014, \citealt{Fab18II}). 
It is unclear whether this preference for a ``simplified'' model is due to statistics, as NGC 7212 has significantly fewer counts than ESO 428$-$G014 ($1280\pm36$ counts compared to $6983\pm84$ counts in their respective 8.0\arcsec~circular regions), or due to confusion, as we are averaging over a bigger physical region in NGC 7212 ($1.5\arcsec=834$~pc) compared to ESO 428$-$G014 ($1.0\arcsec=113$~pc). The presence of dust lanes and disturbed irregular shape (compared to the more disky shape of ESO 428$-$G014) observed in NGC 7212 may also play a role (e.g., \citealt{Mun07}). 

Since our spectral models are consistent across all three regions of interest, we can begin to trace the full physical picture of NGC 7212. 
The low temperature ($\sim$0.8 keV) thermal emission in the inner $1.5\arcsec$~nuclear and $1.5\arcsec-8.0\arcsec$~annulus regions may correspond to shocks with velocities of $v\sim$850 km s$^{-1}$ (assuming $T_{\text{shock}}=3\mu v_{\text{shock}}^2 / 16 k$; where $\mu$ is the mean molecular mass of a fully ionized gas, and $k$ is the Boltzmann constant; e.g., \citealt{Wan14,Fab18II}).

In the annular region ($1.5\arcsec-8.0\arcsec$), the best-fit mixed model prefers a combination of a higher temperature and lower temperature thermal model. This low temperature thermal component is consistent with what is observed in the nuclear region, suggesting similar energetic origins.  
The higher temperature component ($\sim$6.8 keV) is isolated to the annular region and corresponds to velocities near $v\sim2400$ km s$^{-1}$, on order of what is expected from [\ion{O}{3}] velocities in the inner nuclear region (e.g., \citealt{Kra00,Kra08}). Shocks at these velocities have been observed on extended scales in CT AGN (e.g., \citealt{Fis13}) and have been associated with starburst-driven winds (e.g., NGC 6240, \citealt{Wan14}). 

Optical ground-based observations and emission line diagnostics (e.g., BPT diagrams; \citealt{BPT81}), from \citet{Con17} (see also \citealt{Con12}) suggest that the ionization mechanism for NGC 7212 is a combination of photoionization and shocks, which is qualitatively consistent with our results. 
Recently, \citet{Ter16} suggests that NGC 7212 is not likely affected by fast shocks based on near-IR observations. In our $1.5\arcsec-8.0\arcsec$ annular region, however, we find a thermal component that may be associated with fast shocks. This is not necessarily inconsistent with the \citet{Ter16} results, since nonradiative, fast shocks would not be observed in the optical/IR.

The photoionization parameters from our best fits also provide constraints on the presence of highly ionized outflows, such as warm absorbers (WAs), in NGC 7212. WAs typically have velocities that range in the thousands of km s$^{-1}$, column densities around $\text{N}_\text{H}=10^{20-21}$ cm$^{-2}$ (e.g., \citealt{Ara13,Fis13}).
These outflows typically originate from the central continuum in the inner 100 parsec (e.g., \citealt{Kro07,Ara15}), however, \citet{Ara18} find that 12\% of quasar outflows are found at distances larger than 1 kiloparsec. 
In many cases, AGN with WAs exhibit bi-conical outflows reflected in their photoionization parameters (e.g., \citealt{And10,Fab18II}).

In the inner nuclear region of NGC 7212, where outflows are typically located, the ionization parameters we find ($\log\text{U}1=1.25)$ and $\log\text{U}1=-1.50)$) are similar to what has been observed in CT AGN before. However, the column densities in this region are too high, and the velocities derived are too low, compared to what is typically seen in WAs. 
In comparison, the $1.5\arcsec-8.0\arcsec$ annular region contains a thermal component with high $\sim$2400 km s$^{-1}$ velocities, and column densities that are more consistent with WAs. 
While we cannot definitely confirm the presence of WAs in NGC 7212, the extended region is a possible host to these high velocity, highly ionized outflows. 

\subsubsection{Radial Profiles and FWHM}

As described in Section \ref{sec:rp}, we see extended emission in the soft and hard X-rays in both the cone (south$-$north) and cross-cone (west$-$east) direction. It is possible to measure the extent of this X-ray emission out to kiloparsec scales by extracting and fitting the radial profiles of the emission for different energy bands.
Looking at the extended region annulus, we find significant extended emission compared to the \textit{Chandra} PSF for each energy bin: for example, in the $6.1 - 6.5$ keV band where we expect to see Fe K$\alpha$, we find an excess of $15.2 \pm 3.9$ counts. 

\begin{figure}[!t]
\begin{center}
\resizebox{75mm}{!}{\includegraphics{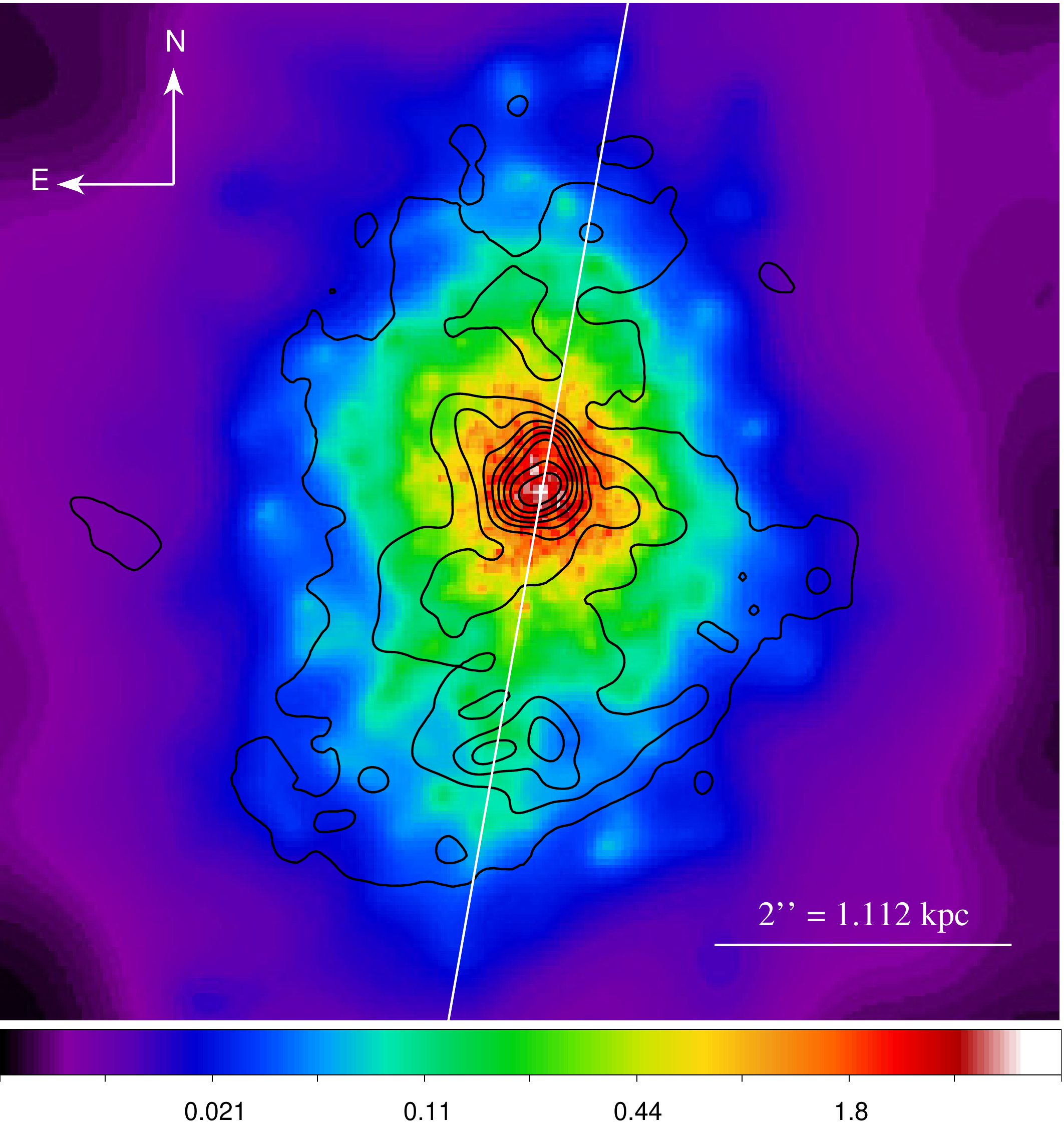}} \\
\caption{Optical contours (black) corresponding to HST HRC F330W observations (\citealt{Sch03}) overlaid on the nucleus of the merged $0.3-7.0$ keV \textit{Chandra} ACIS image of NGC 7212 (adaptively smoothed on image pixel $= 1/8$ ACIS pixel; $0.5-15$ pixel scales, 5 counts under kernel, 30 iterations) The image contours are logarithmic with colors corresponding to the number of counts per image pixel. Also indicated is the orientation of the ENLR (P.A. $=170\degree$, white).
The optical contours are logarithmic with X-ray colors corresponding to number of counts per image pixel. \label{fig:img:opt}}
\end{center}
\end{figure}

The extended region ($1.0-8.0$\arcsec~ annulus) contributes to 20.5\% of the total observed counts, where 17.1\% of this emission is in the soft X-rays. Breaking this down further into the cone and cross-cone regions, we find 14.4\% of the total observed counts originate from the annulus in the south$-$north cone.
The south$-$north cone, in addition to containing the majority of the extended X-ray emission, also encompasses the optical ionization cone and ENLR (Figure \ref{fig:img:opt}; \citealt{Sch03}; see also \citealt{Cra11,Con17}) located at position angle, P.A. $ = 170$\degree. Other works have also uncovered extended emission in the optical and IR that is consistent along this cone axis (e.g., \citealt{Her15,Asm16,Mul18}).

We find a trend in the extended emission with energy in both the cone and cross-cone directions at 1\% of the surface brightness such that the emission is more extended at soft X-rays. For the south$-$north cone, in particular, the soft X-rays are significantly extended on kiloparsec scales (average width $\sim$3.7 kpc).

Our observations also show strong extended emission in the cross-cone direction where we expect significant obscuration from the torus in the ``standard'' model. The origin of the diffuse emission is likely ionizing radiation from the active nucleus, e.g., from the corona, propagating to large scales via interactions with the interstellar medium (ISM). \citet{Geo19} fit a torus model to \textit{NuSTAR} observations of NGC 7212 using MYTorus\footnote{http://mytorus.com/} and found the best-fit parameters $\text{N}_\text{H}=1.1\times10^{24}$ cm$^{-2}$ (consistent with \citealt{Mar18}), $\text{E}_\text{cut}>56$, $k\text{T}=41$ keV.

The presence of a clumpy torus in NGC 7212 could explain this excess emission, such that transmission occurs along the plane of the obscuring material. If we assume a torus geometry for the absorber with an opening angle of $90\degree$ in the south$-$north cone, we can estimate the transmission in the cross-cone direction. With these simple assumptions, we calculate that the volume of cross-cone region is $\sim$$2.5\times$ the volume of the cone. 
From Table \ref{tbl:cts}, we find that the cross-cone region contains $\sim$$37\%$ more counts over the \textit{Chandra} PSF than the south$-$north cone for energies $>1.5$keV. Thus, the transmission in the cross-cone direction is $\sim$$15\%$ of the cone direction.
For comparison, this is higher than the cross-cone transmission estimated for ESO 428$-$G014 ($\sim$$10\%$), but could be due to the weaker azimuthal dependence of NGC 7212, and may be explained by inclination effects.

Alternatively, this extended emission could be related to the compact double radio source observed in NGC 7212 (e.g., \citealt{Tra95I,Tra95II,Fal98}) that can be attributed to the presence of a jet. 
Recent relativistic hydrodynamical simulations of jets propagating through molecular disks (e.g., IC 5063; \citealt{Muk18}) have modeled the presence of warm and hot emission caused by jet$-$cold disk interactions. 
In these models, regions with gas temperatures $\sim$$10^{7}$K may be found surrounding both cooler gas in the nucleus as well as expelled on large scales via filamentary winds. This is in line with our best fit spectral models that fit a thermal component with $kT \sim 0.9$ keV ($\sim$$10^7$ K) in both the $1.5\arcsec$~nuclear and $1.5\arcsec-8.0\arcsec$~annular region.

The excess we observe is not oriented solely along the radio source axis (P.A. $ = -7^\degree$; \citealt{Fal98}). However, the presence of hot gas in the cross-cone region may be explained by recent simulations that predict the presence of a hot cocoon with gas temperatures $\sim$$10^{8-9}$ K surrounding the nucleus due to jet-ISM interactions. 
We find a thermal component in the $1.5\arcsec-8.0\arcsec$~annular region with $kT \sim 7.0$ keV ($\sim$$10^8$ K) that is not found in the $1.5\arcsec$~nuclear region which may indicate the presence of one of these hot cocoons. 

Further evidence supporting jet-ISM interactions in NGC 7212 is reported by \citet{Con17}. However, we cannot rule out supernova heating in the $1.5\arcsec$~nuclear region or the $1.5\arcsec-8.0\arcsec$~annular region. 
The thermal component ($kT \sim 0.9$ keV) in the $1.5\arcsec$~nuclear region corresponds to a cooling time of $\sim$$10^6$ yr. Given an energy content of $\sim$$10^{56}$ erg, the supernova rate required to support this thermal energy is $\sim$$0.1$ yr$^{-1}$.
In the $1.5\arcsec-8.0\arcsec$~annular region, assuming the thermal component  $kT \sim 6.84$ keV is the dominant source of heating, the cooling time is $\sim$$10^8$ years.
For an energy content of $\sim$$10^{57}$ erg, the heating could be accounted for with a supernova rate of $\sim$$0.01$ yr$^{-1}$. 

\section{Summary and Conclusions}\label{sec:con}

We have analyzed the spectral properties and spatial extent of the kiloparsec-scale diffuse X-ray emission in NGC 7212 using 149.87 ks of combined \textit{Chandra} observations. 

\begin{enumerate}

\item We find that the extended diffuse emission region ($1.5\arcsec-8.0\arcsec$ $\sim$$0.8-4.5$ kpc annulus) accounts for more than 20\% of the total emission from 0.3 to 7.0 keV. We further break this down into the soft (0.3 - 3.0 keV) and hard (3.0 - 7.0 keV) X-rays, where we find contributions to the total observed emission of $\sim$17\% and $\sim$3\%, respectively. The energy bin surrounding the Fe K$\alpha$ emission line and hard X-ray continuum (6.0 - 7.0 keV) supplies $\sim$2\% of the total observed emission.

\item Breaking the observed emission into discrete energy bands and cone/cross-cone regions, we find significant, up to 3.6 kpc, extended emission at 1\% of the surface brightness. This extended emission is strongly associated with soft X-rays in the south$-$north cone direction (accounting for more than 14\% of the total counts), and is consistent with the observed ENLR region (e.g., \citealt{Cra11,Con17}) and compact double radio source (e.g., \citealt{Tra95I,Tra95II,Fal98}). In the soft X-rays, this $\sim$3.7 kpc extended emission is similar in extent to the $\sim$3.4 kpc extended soft X-ray emission observed in ESO 428$-$G014 (\citealt{Fab18II}), but remains the largest extent to date in the literature. In the hard X-rays, we observe emission at 1\% of the total surface brightness up to 2.7 kpc (similar to the 2.8 kpc extent in ESO 428$-$G014).

\item The detected emission along the cross-cone direction raises doubts about the standard AGN model in which the hard X-rays originate from the excitation of a uniform obscuring torus. The presence of hard X-rays on observed kiloparsec scales (e.g., for $5.0-7.0$ keV the extent at 1\% surface brightness is $\sim$2.13 kpc) could imply an interior clumpy torus structure (e.g., \citealt{Nen08}) that allows for the transmission of radiation on kiloparsec scales. In the event of a clumpy torus, we estimate that transmission in the cross-cone direction is $15\%$ of the cone direction. This is higher than the cross-cone transmission of $\sim$$10\%$ calculated for ESO 428$-$G014 (\citealt{Fab18II}).

\item We extract the spectrum for three different regions: an 8.0\arcsec~(4.448 kpc) circular region, 1.5\arcsec~(0.834 kpc) nuclear region, and 1.5\arcsec$-$8.0\arcsec~annulus. For each region we fit a reflection model (PEXRAV) plus gaussian emission lines, incrementally adding lines until we obtain a good representation of the spectrum. For each region, the emission line model contains complex, blended emission lines in the soft X-rays below 1.2 keV likely made of \ion{O}{7}, \ion{Ne}{9}, and a variety of Fe lines. Strong individual lines are also found above 1.2 keV, including \ion{Mg}{11}, \ion{Si}{8}, S K$\alpha$, and Fe K$\alpha$ near $\sim$6.4 keV. The presence of these emission lines is consistent with lines found in other AGN (e.g., \citealt{Kos15,Fab18II,Mak19}). 

\item In our spectral line fits, we also discover three significant emission lines in the hard X-rays that are not typically observed in CT AGN. The emission lines at $\sim$2.9 keV and $\sim$3.6 keV are tentatively identified as species of argon and/or calcium. The emission at $\sim$5.2 keV is puzzling and not identified.

\item We also fit the spectra extracted for our three regions utilizing a combination of both photoionization and thermal physical models. In the inner $1.5\arcsec$ nuclear region, we find the spectrum is best fit by a minimum of one photoionization and one thermal model component (although two photoionization model components are preferred). Similarly, the $1.0\arcsec-8.0\arcsec$~annulus region is best fit by at least one of each model component, but a second thermal component is preferred. Combining these two regions, we find the best-fit spectral model for the $8.0\arcsec$~circular region is given as combination of two-photoionization and two-thermal models with model parameters that mirror the individual subregions. The derived parameters in all three regions are consistent with typical ISM densities.

\item We find the ionization parameters in each region of interest are consistent with those found in highly ionized outflows (WAs). However, the typical velocities and column densities of WAs are more in line with the parameters derived in the $1.0\arcsec-8.0\arcsec$~annulus region, rather than near the central source. 

\item We find that the best-fit thermal spectral components for NGC 7212 may be equally well explained by shocks, jet-ISM interactions, and supernova heating. 

\begin{enumerate}
\item In the inner $1.5\arcsec$ nuclear region and $1.0\arcsec-8.0\arcsec$~annular region, the temperature at $kT \sim 0.8$ keV is consistent with $\sim$850 km s$^{-1}$ shocks. 
The additional thermal component in the $1.0\arcsec-8.0\arcsec$~annulus region ($k\text{T}=6.84$ keV) corresponds to shocks near $\sim$2400 km s$^{-1}$, and is consistent with previous observations of CT AGN (\citealt{Fis13}). 

\item The warm thermal component ($kT \sim 0.8$ keV) can likewise be explained by jet-ISM interactions in which warm gas is found surrounding cool gas in the nucleus and expelled on large scales via filament winds. Similarly, simulations of jet-ISM interactions predict a hot cocoon around the nuclear region that is consistent with the hot thermal component we observe in the annular region of NGC 7212 (e.g., \citealt{Muk18}).

\item We are unable to rule out supernova heating as the origin of this thermal component in both the $1.5\arcsec$ nuclear region and $1.0\arcsec-8.0\arcsec$~annular region. 
For the $1.5\arcsec$ nuclear region where $kT \sim 0.8$ keV (cooling time $\sim$$10^6$ years; E$_{\text{th}}\sim$$10^{56}$ erg), the heating could be accomplished with $\sim$0.1 supernova per year. Assuming the hot thermal component ($kT \sim 7$ keV; cooling time $\sim$$10^8$ years; E$_{\text{th}}\sim$$10^{57}$ erg) dominates in the $1.0\arcsec-8.0\arcsec$~annular region, the supernova rate drops to $\sim$0.01 yr$^{-1}$.
\end{enumerate}

\item Compared to other CT AGN with extended X-ray emission, the most comparable of which is ESO 428$-$G014 (\citealt{Fab18II}), we find that NGC 7212 requires a less complex multicomponent spectral model. These differences may be purely due to statistics (our observations of NGC 7212 have combined $\sim$1300 counts, compared to $\sim$7000 counts for ESO 428$-$G014), or confusion due to spatial scale (for NGC 7212, $1\arcsec=556$ pc, compared to $1\arcsec=113$ pc for ESO 428$-$G014).

\end{enumerate}

This work demonstrates the advantages of deep \textit{Chandra} observations for recovering statistically significant spatial and spectral information about an exciting class of CT AGN with observed diffuse hard X-ray emission on kiloparsec scales. High-resolution observations of extended emission sources, such as NGC 7212, can recover important information about the AGN and surrounding ISM, which can be used to test the AGN standard model by analyzing the morphology of the torus, and provide new insights into gas dynamics and AGN feedback mechanisms. 
As we plan for the next generation of great observatories (e.g., \textit{Lynx}), an emphasis on high spatial resolution and sensitive instruments will play a crucial role in the future of AGN studies.

\acknowledgments

We thank E. Bulbul and A. Foster for their assistance in identifying challenging X-ray emission line features, and P. Plucinsky for a useful discussion about the \textit{Chandra} ACIS background. We also thank the referee for constructive comments and suggestions that improved this paper. This work makes use of data from the \textit{Chandra} data archive, and the NASA-IPAC Extragalactic Database (NED). The analysis makes use of CIAO and Sherpa, developed by the \textit{Chandra} X-ray Center; SAOImage ds9; XSPEC, developed by HEASARC at NASA-GSFC; and the Astrophysics Data System (ADS). This work was supported by the \textit{Chandra} Guest Observer program, grant no. GO8-19074X (PI: Fabbiano).


\bibliography{v1.8}


\end{document}